\documentclass{article}

\usepackage{microtype}
\usepackage{graphicx}
\usepackage{subfigure}
\usepackage{booktabs} 

\usepackage{hyperref}

\usepackage[accepted]{icml2024}

\usepackage{amsmath}
\usepackage{amssymb}
\usepackage{mathtools}
\usepackage{amsthm}

\usepackage[capitalize,noabbrev]{cleveref}

\usepackage{amsfonts}
\usepackage{graphicx}
\usepackage{textcomp}
\usepackage{xcolor}
\usepackage{algorithm}
\usepackage{subcaption}
\usepackage{tikz}
\usepackage{filecontents}
\usepackage{listings}

\theoremstyle{plain}
\newtheorem{theorem}{Theorem}[section]

\theoremstyle{definition}
\newtheorem{definition}[theorem]{Definition}

\theoremstyle{remark}

\begin{document}
	
	\twocolumn[
	\icmltitle{Exploring System-Heterogeneous Federated Learning with Dynamic Model Selection}

\begin{icmlauthorlist}
	\icmlauthor{Dixi Yao}{yyy}
	
	\icmlaffiliation{yyy}{University of Toronto}
	\icmlcorrespondingauthor{Dixi Yao}{dixi.yao@mail.utoronto.ca}
\end{icmlauthorlist}

	\icmlkeywords{Federated Learning, AutoML, System Heterogeneity}

	\vskip 0.3in]
	\printAffiliationsAndNotice{} 
	
	\begin{abstract}
		Federated learning is a distributed learning paradigm in which multiple mobile clients train a global model while keeping data local. These mobile clients can have various available memory and network bandwidth. However, to achieve the best global model performance, how we can utilize available memory and network bandwidth to the maximum remains an open challenge. In this paper, we propose to assign each client a subset of the global model, having different layers and channels on each layer. To realize that, we design a constrained model search process with early stop to improve efficiency of finding the models from such a very large space; and a data-free knowledge distillation mechanism to improve the global model performance when aggregating models of such different structures. For fair and reproducible comparison between different solutions, we develop a new system, which can directly allocate different memory and bandwidth to each client according to memory and bandwidth logs collected on mobile devices. The evaluation shows that our solution can have accuracy increase ranging from 2.43\% to 15.81\% and provide 5\% to 40\% more memory and bandwidth utilization with negligible extra running time, comparing to existing state-of-the-art system-heterogeneous federated learning methods under different available memory and bandwidth, non-i.i.d.~datasets, image and text tasks.
	\end{abstract}

	\section{Introduction}
Federated learning (FL)~\cite{fedavg} is a distributed learning paradigm to train a global model over multiple mobile devices. With the help of mobile devices' edge computation power, network communication and a centralized server, the system is able to keep all users data local and guarantee the privacy. As servers in the cloud are usually powerful, the edge device's computation ability and network bandwidth become the bottleneck of training models to high accuracy efficiently. To address such a bottleneck in federated learning system, large efforts have been put to optimize the edge computation and communication through model compression~\cite{hermes,fedscr,fedskel,fedhq}, adaptive batch sizes~\cite{batch}, scheduling~\cite{yao2021context}, regularization~\cite{feddyn}, and federated neural architecture search~\cite{fedrlnas,fednas,garg2020direct,fedoras,yao2024perfedrlnas} etc.  

To address the issue that different mobile devices can have different memory, bandwidth, etc., system-heterogeneous federated learning methods are proposed to assign models of different sizes or different computation complexities (e.g. Flops) to each mobile device~\cite{fjord,heterofl, fedrolex, feddropout,fedhq,anycostfl}. Previous solutions were based on the assumption that each device's running status (e.g. network connection, memory usage, etc.) would not change during the federated learning process. But in the real world, such an assumption will no longer hold~\cite{yao2021context,guo2021mistify,wang20context,huang2020clio,almeida2022dyno}. 

However, in practical applications, there is no guarantee that every time the same device checks in, its available memory or bandwidth will remain the same. For example, during the training process, the memory resources on a mobile phone will be constrained if a user leaves models training in the background and later the application will be closed. On the other hand, if a device is always of low memory resources, it will never check in.

Another problem with existing system heterogeneity work is that in their evaluation, they do not assign models to each client directly based on available memory. Prior studies on system-heterogeneity have relied on a simulation approach that assigns a fixed number of mobile devices with a designated model complexity during each communication round. However, such a type of simulation cannot show performance in practical use cases.

Different from previous evaluation processes, we develop a system-heterogeneous federated learning system based on \textsc{Plato}\cite{plato}. Our system can simulate the clients' available resources (memory, bandwidth, etc.) with logs collected on physical devices (e.g.~network speed, memory logs). With our system, we can have reproducible evaluation over different algorithms fairly with the same configuration files. We argue that we should perform experiments directly over changing resources such as memory and bandwidth available on the devices, and that the selection of models assigned to clients is based on their real-time available memory and bandwidth.

On running our system, we find that existing solutions may achieve sub-optimal results in certain settings. We observe that existing solutions adopted the same channel pruning rate on each layer. The heterogeneity in models is only based on one dimension of different channel numbers. This leads to the consequence that we have a limited range of selecting different model structures for each client. Hence, each client may not receive the model fitting into its memory or bandwidth the best, resulting in sub-optimal performance.

Motivated by these empirical observations, we propose to directly assign models of different architectures to mobile devices according to their available memory and bandwidth in each round. These models not only vary in channel numbers on each layer but also vary in the number of layers. With more choices of different architectures and finer grained selection of models, our method can cover a much wider range of cases over the mobile devices in federated learning.  With the large flexibility of the model architectures brought on, the whole search space is also very large. Hence, we further propose a constrained search method, gradually expanding search space through random search.

During the model aggregation stage on the server, as models are of different architectures, we further develop an original federated in-place distillation method to improve the performance. Different from conventional knowledge distillation~\cite{kd} methods, we do not need any extra data (neither generated nor public datasets) to conduct distillation. Instead, we conduct knowledge distillation between subsets of the global model and the global model itself on the server. The proposed distillation module can also be applied to previous system-heterogeneous methods.

With our system, we revisit various federated learning settings and train different models including convolution neural networks and transformers over i.i.d.~and non-i.i.d.~datasets. In the evaluation, we test the system with bandwidth logs of running HTTP/2 applications. It turns out that our models have a better utilization of bandwidth and memory over clients up to 40\%. It is shown that we can improve the accuracy by 2.43\% to 15.81\% among image and text classification tasks. 
	\section{Related Work and Motivation}
\subsection{System Heterogeneity in Federated Learning}
\label{subsect:form}
In the context of federated learning (FL), we have $N$ clients on mobile devices. Each client has its own local data $\mathcal{D}_i,i\in[N]$. The data on these clients can be either i.i.d.~or non i.i.d. The server aims to train a global model $w$ which can be utilized by all these clients. 

In a practical FL system, each mobile device can have particular resource constraints and not all of them are able to run a very large global model on their devices. We call such differences in resource budgets between these clients \emph{system heterogeneity}. 
Moreover, such resource budgets can also change frequently along the federated learning when users are running applications~\cite{yao2021context,guo2021mistify,wang20context,huang2020clio,almeida2022dyno}. Memory usage and bandwidth are two common resource budgets in mobile computing~\cite{yao2021context}. We do not consider energy here as users will probably not frequently change energy mode during the few hours learning process.

\begin{figure}
	\centering
	\includegraphics[width=0.95\linewidth]{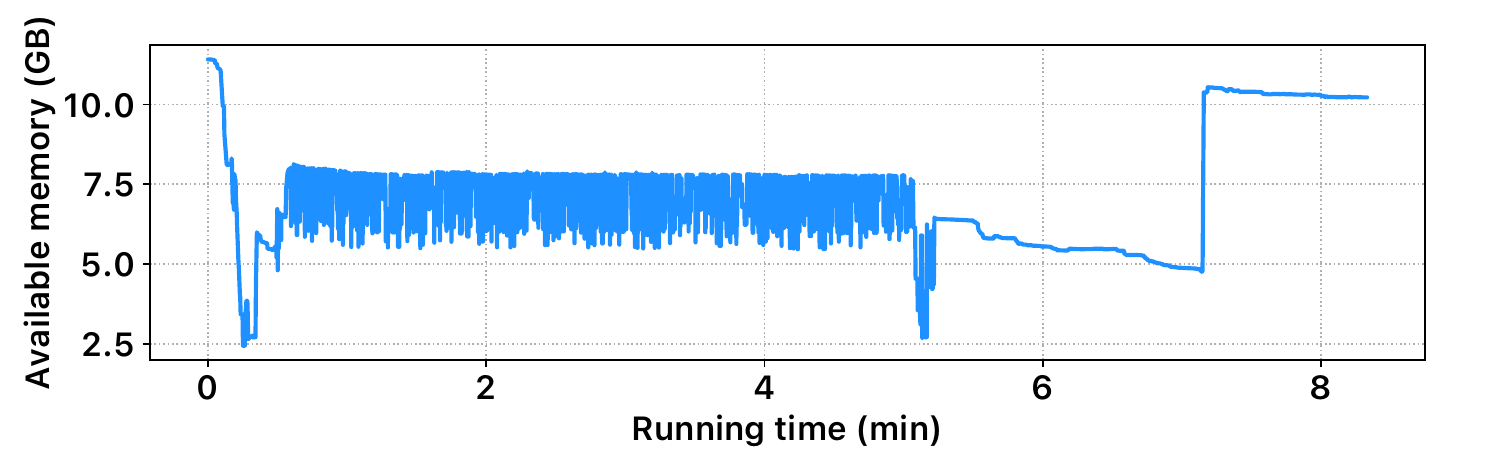}\hfill
	\includegraphics[width=0.95\linewidth]{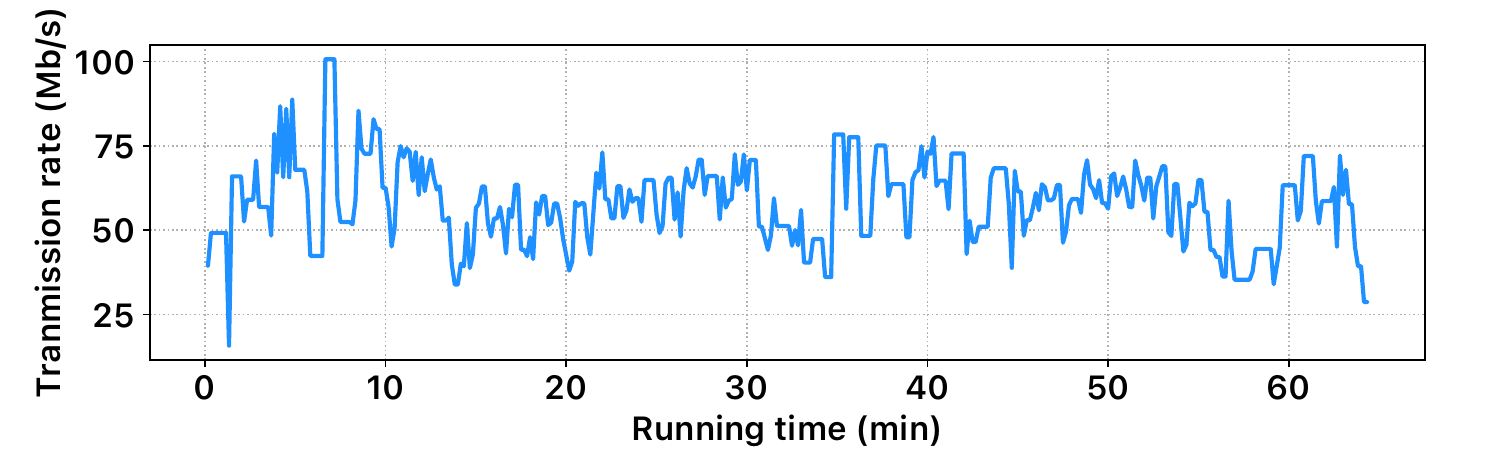}
	\caption{The changing of available memory and transmission rate when running the applications. }
	\label{fig:flucbudget}
\end{figure}

As existing edge devices are equipped with GPUs (Jetson TX1, iPhone, MacBook, etc.~), we mainly consider GPU memory. As shown in \cref{fig:flucbudget}, we measure the available memory on the Apple MacBook Pro with M1 chip while running an image generation application ControlNet~\cite{zhang2023adding}. The available memory will fluctuate during the running of the application, from 2GB to 10GB. We also consider the transmission rate, as in a mobile environment, communication is important. Many mobile devices has the bandwidth~\cite{heterofl, fedrolex, feddropout, hermes,fedrlnas} limitation and cannot transmit so many model parameters at the same time. The transmission rate or the network speed will change frequently~\cite{yao2021context,guo2021mistify}. In \cref{fig:flucbudget}, we visualize the bandwidth logs recorded on the MacBook using a public Wi-Fi of 5GHz bandwidth. The network transmission rate can frequently change on user-end devices. Therefore, we need a system that can handle such heterogeneity with fluctuating resource budgets. 

\subsection{More Flexibility in Different Model Architectures Can Provide Better Performance}
Though each client device is not able to run the largest model which has the best accuracy, we still want to learn a global model having as high accuracy as possible. Conventional methods in efficient FL \cite{hermes,fedscr,fedskel,fednasbig,fedhq,batch} did not base their optimization over actual resource constraints on each client, making the assigned models still possibly fail to run on the devices. Hence, a series of system-heterogeneous federated learning methods are proposed. HeteroFL~\cite{heterofl} leveraged channel pruning to assign models of different channel widths. FjORD~\cite{fjord} further proposed an ordered dropout knowledge distillation module. FedDropout and Split-Mix FL~\cite{feddropout,hongefficient} leveraged similar idea but they selected the pruned channels randomly. \citeauthor{dun2023efficient}~\cite{dun2023efficient} expanded the idea of randomly selecting pruned channels through Dropout layers and applied this idea into asynchronous scenario. FedRolex~\cite{fedrolex} further introduced a rolling scheme to the pruning process, applying a more balanced node selection policy. AnyCostFL~\cite{anycostfl} introduced a mechanism that in each communication round, channels are first sorted on basis of their importance and each client choose top-$k$ important channels. However, all these previous methods choose to only prune the channel of each client model at the same rate for all layers.

Nevertheless, assigning clients with models of different model architectures can bring much more flexibility. We conduct a toy experiment. We have 100 clients and 10 out of them are selected in each round. We construct a non-i.i.d.~CIFAR10~\cite{cifar10} dataset with Dirichlet distribution (parameter $\alpha=0.1$). The local epoch in each round is 5. If a ResNet152 model is pruned by $62\%$ channels on each layer to have a similar size ResNet18 model, the resulting pruned model only achieves \textbf{68.73\%} accuracy, while ResNet18 can reach \textbf{71.91\%}. The original ResNet152 can achieve {\textbf{74.51\%}}. As a result, leveraging the smallest models on all clients or channel pruning are neither good options. Assigning models of different structures to each client is intuitively better. 

Federated neural architecture search~\cite{yao2024perfedrlnas,fedrlnas,fednas,fednasbig,fedoras} is a more general method of model search in federated learning. Neural architecture search (NAS) is a technique for automating the design of artificial neural networks. As a result, guided by this design idea, existing methods are either targeted for personalized federated learning~\cite{yao2024perfedrlnas,fedoras} or need to transmit and compute large supernets on the clients~\cite{fednas,fednasbig,fedrlnas}. Such objectives make existing federated neural architecture search methods fail to meet the requirements of finding suitable models with given constrained memory and network bandwidth.

\subsection{In-place Distillation}
In centralized neural architecture search, employing in-place distillation enhances the supernet's performance significantly~\cite{simnas,bignas}. In one-stage NAS supernet training, multiple sub-networks are sampled at each training step. In the centralized NAS approach, the entire supernet is initially trained directly on the dataset using real labels. Subsequently, the sampled sub-networks are trained using logits from the supernet rather than real labels. The NAS then aggregates all sampled networks into the supernet. In summary, denoting the supernet as $w$, the sampled sub-networks as $w_s$ (representing this set), and the centralized dataset as $\mathcal{D}_c$, the supernet is trained using a specific loss function:
\begin{equation}
	\label{eqa:kd}
	\mathcal{L}=\mathcal{L}_{\mathcal{D}_c}(w)+\gamma \mathbb{E}_{w_s}\mathcal{L}_{\rm KD}([w,w_s];w_{t-1})
\end{equation}
where $w_{t-1}$ means the weights of the supernet in the last iteration and $\mathcal{L}_{\rm KD}$ is the distillation loss, e.g.~Kullback–Leibler divergence. However, we can not directly leverage such a process into the model search process in the federated learning setting. Because the data is safely kept and distributed over the clients, we can neither conduct the first step of training the supernet with a centralized dataset nor the second step of training subnets with soft labels. FjORD~\cite{fjord} proposed a self-distillation process, conducted on the clients. However, as data is probably non-i.i.d.~in FL, such an operation can misguide the distilled model overfit on local datasets and cause severe performance degradation. Besides, carrying out distillation process on the clients will add on more overhead on clients.

Other solutions for system-heterogeneous federated learning based on knowledge distillation were also proposed. FedMD~\cite{fedmd} can aggregate client models of different user self-designed structures. FedDF~\cite{feddf} aggregates the client models and the global model through knowledge distillation over unlabeled public data. Fed-ET~\cite{fedet} proposed a weight consensus distillation scheme with diversity regularization. \citeauthor{zhu21data}.~\cite{zhu21data} proposed to train a data generator on the server to generate data for knowledge distillation. However, these methods rely on public or generated data. Furthermore, in these methods, client model weights are partially or entirely transmitted to the server. Such design are incompatible with secure aggregation protocols violating the initial design of FL. 
	\section{Dynamic Model Pruning for System Heterogeneous FL}
Though it is intuitive that extending HeteroFL, FedRolex, etc.~from simple channel pruning to searching models of different architectures can bring more performance improvement, it is non-trivial to put it into practical use cases. We show the overall workflow in \cref{fig:workflow}. The algorithm version is in \cref{appendix:overallalg}. The server holds the global model. In each communication round, the server will sample single models of different architectures with \textbf{resource constrained model search} and then send them to each client participated in this round. 
Each client will conduct local training with assigned models and their own data for several local epochs. This is the same as the process in FedAvg. After that, they upload the trained model weights as well as the information about available memory and bandwidth to the server. The server will first aggregate models of different architectures into the global model. 
We will conduct several iterations of \textbf{federated in-place distillation} on the server to improve the global model performance. The system will then run into the next round.

\begin{figure}[tb]
	\centering
	\includegraphics[width=0.8\linewidth]{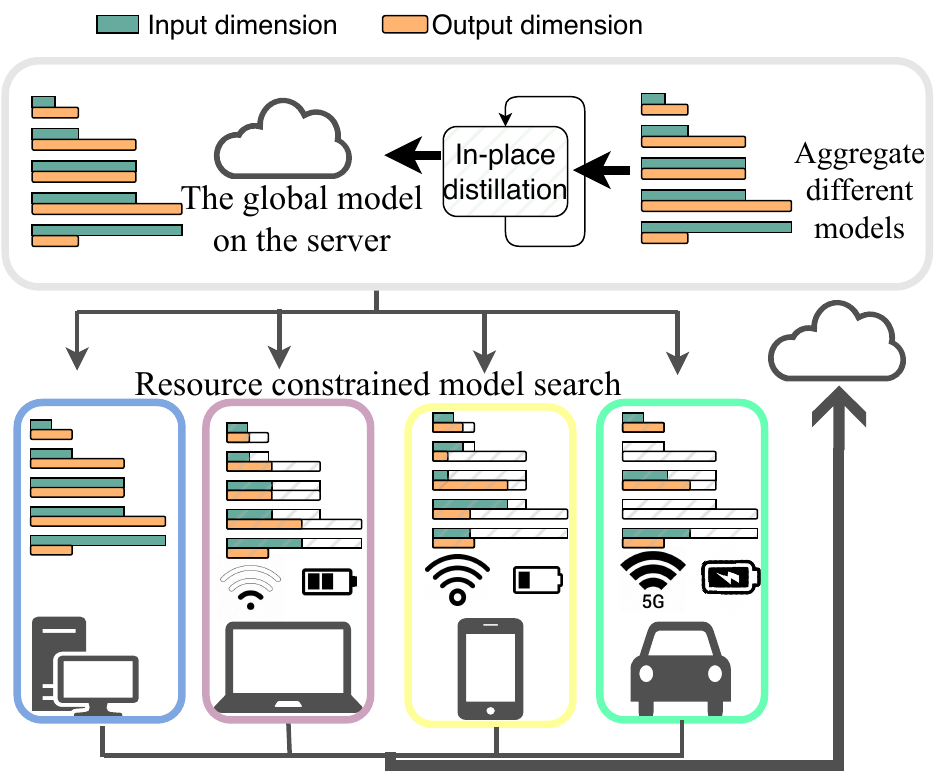}
	\caption{The overall workflow of our method to train the global model in system-heterogeneous FL.}
	\label{fig:workflow}
\end{figure}
\subsection{Global Model Design and Aggregation}
We build the global model into several search dimensions including depths, widths, etc. For each sampled client model and for each layer, a feature dimension will be searched. If it is $0$, it means such a layer is not selected. 

\begin{definition}
	\label{def:supernet}
	(The global model and search space) Given a global model $w$ with $d$ layers. We have a search space $(w,S)$ where $S=\{s_1,\cdots, s_m\}$, a ratio set for the search options on each layer ($m\geq1)$. Each search option defines the output dimension of the layers .
	$S$ has least one element $s_m=1$ representing no change in the original operation.
\end{definition}

The $S$ in the search space can either be a continuous space (e.g. channel pruning rate in (0,1]) or a discrete space.

\begin{definition}
	\label{def:subnet}
	(The sampled client model) With a given search space $(w,S)$. We can represent a subnet $w_i$ as $w\otimes V$ where $V=(v_1,\cdots,v_d)$ and $d$ is the number of the layers of $w$, except for the last layer $v_i\in S, \forall i\in[d]$.
\end{definition}
As a result, we can use $(w,(1,\cdots,1))$ to represent the global model. 
The previous methods of only one dimension of channel pruning rate can be viewed a special case of our problem where the search space only has $m$ choices: from $(w,(s_1,\cdots,s_1))$ to $(w,(s_m,\cdots s_m))$.

\begin{figure}[tb]
	\centering
	\includegraphics[width=\linewidth]{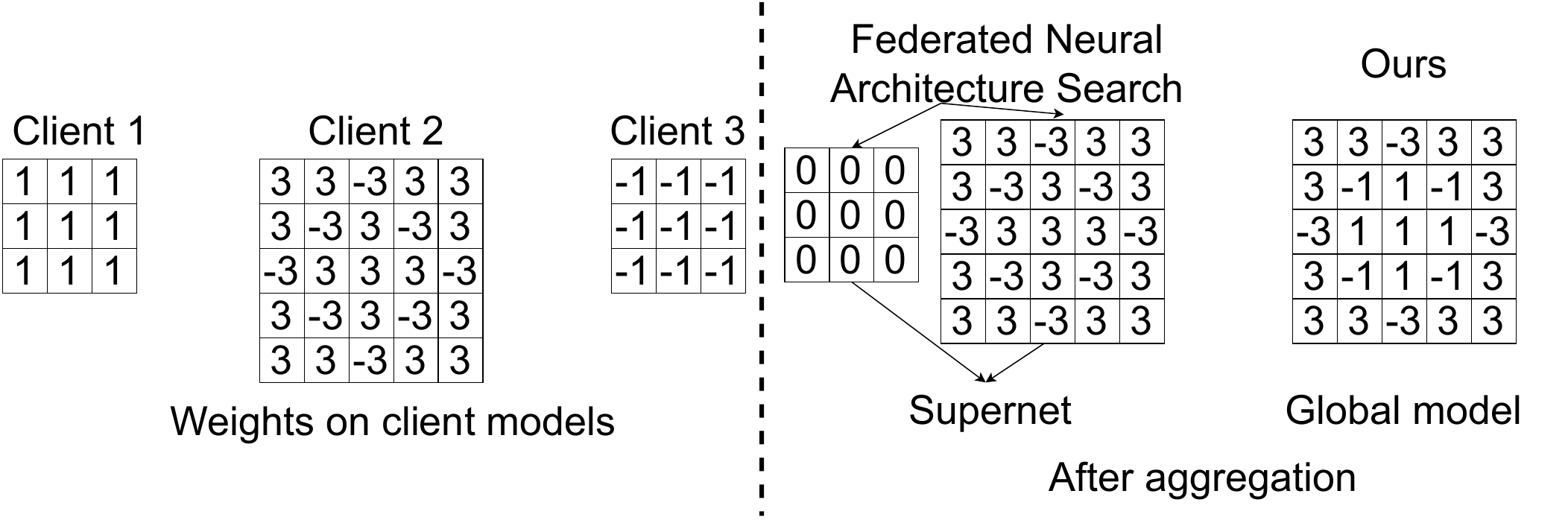}
	\caption{The illustration of our aggregation method of parameters from different clients. The example shows one convolution layer with different kernel sizes on the clients.}
	\label{fig:aggregation}
\end{figure}

We construct the client model according to our representation of $(w,V_i)$. With the given search space, we are able to sample a subnet for each client in each communication round. The next problem is aggregating models of different architectures to the global model after the client uploads them. However, since we have models of different architectures, we cannot directly add them and make the division. To resolve this, instead of aggregating on the operation level like in previous work~\cite{fedrlnas,fedoras}, we do average over the element-wise level. For example, as shown in \cref{fig:aggregation}, in previous methods, $3 \times 3$ convolution layer and $5 \times5 $ convolution layer will be treated as two operations and their parameters will be aggregated separately. While in our method, the parameters in $3 \times 3$ will be aggregated together with parameters in $5\times 5$ kernel.

\subsection{Resource Constrained Model Search}
During the sampling phase, we will sample an architecture from the search space which is the largest model satisfying the given resource condition of each client. With the larger flexibility brought by the large architecture search space, a challenge is that it is impossible to list all possible architectures ($m^d$ possible architectures in all) and search for one which meets the requirements. 

As a result, we gradually expand a sampling pool through the training process instead of listing all possible architectures out. At the very beginning, the sampling pool $\mathcal{P}$ only has the largest model (the global model) and the smallest sub-net. In each step of sampling a client model, we first traverse the sampling pool and find the largest architecture satisfying the requirement. By meaning largest, it means the architecture has the most parameters. To expand the sampling pool and have more candidates, we will have probability of $\epsilon$ to adopt the searched architecture and a probability of $1-\epsilon$ to random search a new architecture from the search space and add it to the sampling pool. The random search will be repeated by maximum of $T_{\rm MAX}$ times each client sampling. In the worst cases, the smallest model will be chosen for the current sampling requirements. 

Our random search scheme ensures that all parameters have an equal opportunity to be selected. Especially, each layer individually has the chance to be selected and the all parameters on that layer will be trained. In FedRolex~\cite{fedrolex}, rolling mechanism is introduced to perform a more balanced scheme of choosing parameters. In our proposed solution, we can address this concern regarding the unbalanced updates of parameters.

\subsection{Federated In-place Distillation}

\begin{figure}[tb]
	\centering
	\includegraphics[width=0.9\linewidth]{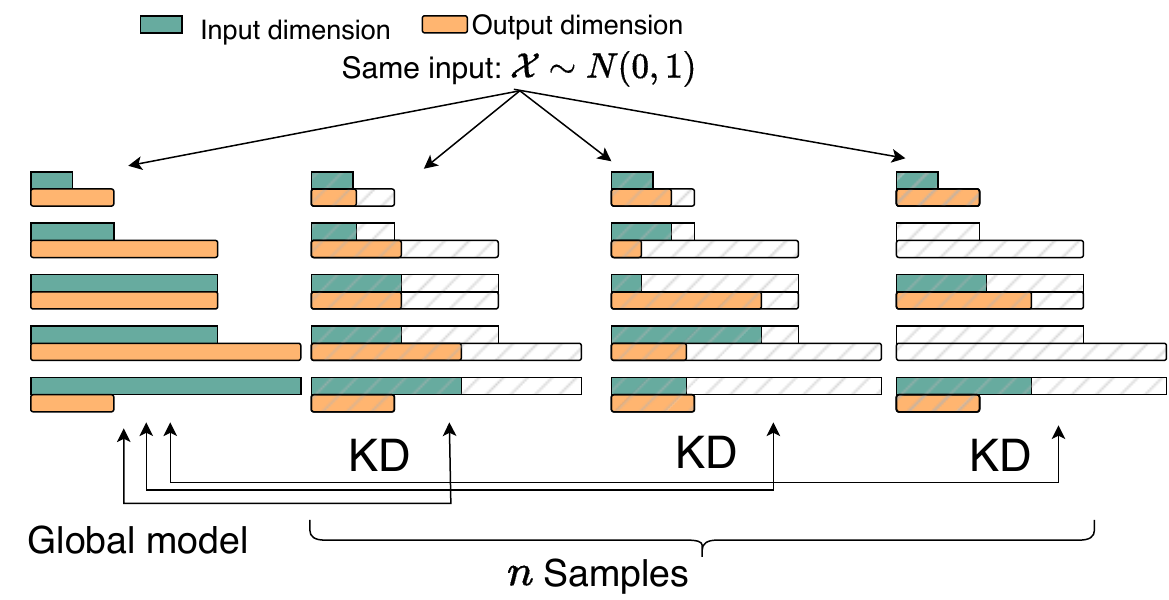}
	\caption{The process of in-place distillation on the server. Notably these $n$ new architectures are sampled from $\mathcal{P}$.}
	\label{fig:skd}
\end{figure}

To further the performance of the global model which is trained through aggregation of models of different structures, we propose an in-place distillation process specific to federated learning. Different from previous distillation method, we do not need public or generated data but can incorporate in-place distillation process during federated learning. The process of in-place distillation is shown in  \cref{alg:skd}. The derivation of \cref{alg:skd} is in \cref{appendix:distill}. After the global model is aggregated from the client models, on the server we randomly sample $n$ subnets of different architectures from the global model (not from $\mathcal{P}$). To make sure no privacy information of clients is leaked to the server. We let each client normalize its local samples on the local datasets first before federated learning. Then, in each iteration of in-place distillation, we generate $K$ samples from a normal distribution $\mathcal{X}_{\rm KD}\sim\mathcal{N}(0,1)$, where $K$ is the batch size. As shown in \cref{fig:skd}, for each sampled subnet, we train them with the soft labels from the global model using the generated $K$ samples in each iteration. After $T_{\rm SKD}$ iterations of training, we aggregate the weights of these $n$ subnets into the global model. Our in-place distillation module can be applied to other methods that aggregate models of different structures such as HeteroFL.

\subsection{Activation Function with Boundary}
In our search method, apart from different model widths, we also have different model depths and each layer can also be directly skipped. To solve such issue, we add an extra ReLU6 ~\cite{mobilenetv2} out of the activation function in the neural network to give a boundary over the activation.
	 \subsection{Convergence Analysis}
\label{subsec:converge}
Compared to previous solutions, while our expanded search space entails traversing more architectures (from $m-1$ possibilities to $m^d$), this does not adversely affect convergence. We hold our conclusion based on the common assumptions in non-convex optimization in federated learning~\cite{convergence,scaffold}
\begin{theorem}
	\label{theorem:convergence}
	If the convergence rate of FedAvg is $\mathcal{O}(\frac{1}{T})$, the convergence rates of model search based and model-heterogeneous based system-heterogeneous federated learning methods are $\mathcal{O}(\frac{1}{T})$. The proof is in \cref{appendix:convergence}.
\end{theorem}

\begin{algorithm}[tb]
	\caption{Federated in-place distillation}
	\label{alg:skd}
	\begin{algorithmic}[1]
		\STATE {\bfseries Input:} The global model $(w,S)$
		\STATE {\bfseries On Server: }\;
		\STATE Random sample $n$ new subnets from $(w,S)$.\;
		\FOR{iterations $\leftarrow$ 1 to $T_{\rm SKD}$}
		\FOR{$i\leftarrow 1$ to $n$}
		\STATE Generate $\mathcal{X}_{\rm KD} $ with $K$ samples.\;
		\STATE Train the sampled subnets $w_i$ with knowledge distillation from the global model.\;
		\ENDFOR
		\STATE Update the global model $w$ with gradients:
		\begin{equation}
			\label{eqa:skd}
			\nabla w=\frac{1}{n}\Sigma_{i\in[n]}\mathcal{L}_{\rm KD}(w,w_i;\mathcal{X_{\rm KD}})
		\end{equation}
		\ENDFOR
	\end{algorithmic}
\end{algorithm}
	\section{System Design}
We implement our system based on \textsc{Plato}\cite{plato}, an open-source framework for FL. \textsc{Plato} provides interfaces easy of use, which can simulate cross-device scenarios with a lot of devices and various data distribution. Plato is compatible with existing FL frameworks and infrastructures, providing a good development base for our system.
\subsection{System Requirements}
First, the system should \textbf{directly simulate the available resources in real-world cases}. In the simulation process in previous methods such as HeteroFL and FedRolex, they equally assign models of different sizes to clients. For instance, with 10 clients per round, each client receives models of varying complexities. It's assumed that there will be precisely two devices capable of running each complexity level, ensuring a balanced distribution. Additionally,1/5 of all clients are expected to qualify for each of the five designated models in every round.


Second, the system should \textbf{measures directly in metrics evaluating resources (memory, bandwidth) rather than proxy metrics (FLOPS, number of parameters).} Our interfaces facilitate accurate assessment of memory and bandwidth usage. When we implement our proposed algorithm, we search for model structures directly through its constraints.

Thirdly, \text{ensuring reproducibility} is crucial for establishing a fair comparison of methods. With the users given system configurations, we can compare various algorithms on board with the same settings and replicate the experiments.
 
\subsection{System Implementation}
\begin{figure}[tb]
	\begin{minipage}{0.5\linewidth}
		\centering
		\includegraphics[width=0.98\linewidth]{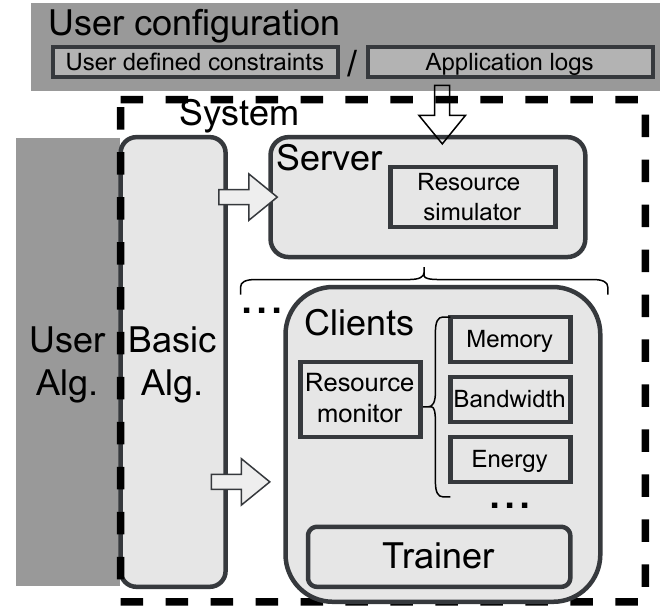}
		\caption{System structure.}
		\label{fig:system}
	\end{minipage}
	\hfill
	\begin{minipage}{0.44\linewidth}
		\begin{lstlisting}[frame=single]
limitation:
  memory:
    binary: false	
    min: 4
    max: 8
  bandwidth:
    log_path: ...
		\end{lstlisting}
		\caption{Configuration.}
		\label{fig:configuration}
	\end{minipage}
\end{figure}

We build the functionalities of our system in four modules \emph{Server}, \emph{Client}, \emph{Trainer}, and \emph{Algorithm} based on \textsc{Plato} interfaces. The overall structure is shown in \cref{fig:system}. We have a server to load in user configuration. The server will correspondingly assign the available memory, bandwidth, etc.~to clients and the clients will monitor the usage of these resources. We implement different system-heterogeneous algorithms in the user algorithm part, on top of the basic algorithm  (FedAvg). In such a way, we can control the only variable: algorithms used, during experiments.

Notably, our design of separating modules can precisely simulate the wall-clock elapsed time in federated learning, for example the time spent during running algorithms and training models in a trainer. It will not count in the overhead spent on simulating the system, for example the time spent in resource simulator and resource monitor. 
	\section{Evaluation}
\subsection{Experiment Settings}
\subsubsection{Setup for Federated Learning}
To have a thorough comparison between different solutions, we choose various federated settings with various number of total clients, number of clients participated in each communication round, local epochs, different datasets and different model families. To have a thorough comparison of different system settings, we have 6 different settings for comparison. Regarding datasets, we choose CIFAR10~\cite{cifar10}, FEMNIST~\cite{leaf}, and Shakespeare~\cite{leaf} as our datasets. We have both i.i.d.~and non-i.i.d.~datasets. For CIFAR10, we use Dirichlet distribution with parameter $\alpha$ to generate non-i.i.d.~datasets. FEMNIST and Shakespeare are originally non-i.i.d.~datasets. More details of client configurations are in \cref{tab:settings}. In setting 1 and 6, we have 3597 and 1128 clients respectively, which is sufficient for showing scalability. On CIFAR10, we train convolution neural networks. On FEMNIST, we train vision image transformers. On Shakespeare, we train transformers. More details about datasets and models are in \cref{sec:datasets}.

The first two experiment settings, including the settings of the systems, hyper-parameters, models adopted and the distribution resource-constraints are the same as settings in non-i.i.d.~cases over the CIFAR10 dataset in HeteroFL~\cite{heterofl} and high data heterogeneity cases over the CIFAR10 in FedRolex~\cite{fedrolex} respectively. The next three settings simulate system heterogeneity in the real-world and the last setting directly uses the devices' logs of network and memory. We will give more details about system configurations in \cref{sysconfiguration}.

Our batch size is fixed at $64$ for all experiments. For resource constrained model search, the early stop parameter $T_{\rm MAX}=5$ and $\epsilon=0.8$. The settings of the rest hyper-parameters are in~\cref{sec:hyper}. For the federated in-place distillation, the global model is trained with Adam optimizer with learning rate 0.001, $T_{\rm SKD}=100$, and $n$ is set the same as number of clients participated in each round. We evaluate the inference accuracy of the global model on the server.

\begin{table}[tb]
	\caption{The client configurations of experiments during FL. CIFAR10 $0.1$ means the dirichlet parameter $\alpha=0.1$.}
	\label{tab:settings}
	\begin{center}
		\small
		\begin{tabular}{ccccl}
			\hline
			Setting & \begin{tabular}[c]{@{}c@{}}Total\\clients\end{tabular} & \begin{tabular}[c]{@{}c@{}}Clients \\participated\\each round\end{tabular} & \begin{tabular}[c]{@{}c@{}} Local\\epochs $\tau$ \end{tabular} & Dataset \\
			\hline
			Setting 1 & 100 & 10 & 5 & CIFAR10 $0.1$ \\
			Setting 2 & 100 & 10 & 1 & CIFAR10 $0.1$ \\
			Setting 3 &50 & 50 & 1 & CIFAR10 i.i.d.~\\ 
			Setting 4 & 100 & 10 & 5 & CIFAR10 $0.1$\\
			Setting 5 &3597 &120 &5& FEMNIST\\
			Setting 6& 1128 &120&1&Shakespeare\\
			\hline
		\end{tabular}
	\end{center}
\end{table}

\subsubsection{System Configurations}
\label{sysconfiguration}

To simulate system heterogeneity in real applications, we sample the memory and bandwidth budgets over each client every communication round.  The distribution of the resource constraints is in  \cref{tab:model}. 
We cover the usual range of available bandwidth and memory on mobile devices~\cite{guo2021mistify,yao2021context,wang20context}. The memory and bandwidth are independently sampled.

To have a fair and thorough comparison, we have four different types of configurations. In setting 1 and 2, the simulation configuration is the same as in the previous work. We exactly sample 1/5 of the clients having the available resources corresponding to the first model complexity, and 1/5 of the clients corresponding to the second and so on.  In setting 3 and 4, memory and network speed are uniformly sampled from the range. In setting 5, the memory is uniformly sampled from the range. We import the devices' logs of transmission rates into the system, using the Wi-Fi logs shown in \cref{fig:flucbudget}. 

In setting 6, we import the logs of transmission rates in real-world applications into the system. We have 4G/LTE bandwidth logs~\cite{ltelog} and Wi-Fi logs. The 4G logs are collected through running HTTP/2 applications over public transportation~\cite{ltelog}. The logs are of a duration of about 1 hour, which can cover the training processes. 
We have memory configurations of 4GB and 8GB, which are common configurations of edge devices~\cite{guo2021mistify,yao2021context}.

\begin{table}[tb]
	\caption{The distribution of resource constraints on mobile devices and the models used. The speed is measure in Mb/s and the GPU memory is measure in GB.}
	\label{tab:model}
	\begin{center}
		\small
		\begin{tabular}{c@{\hspace{5.5pt}}l@{\hspace{5.5pt}}l@{\hspace{5.5pt}}c@{\hspace{5.5pt}}c}
			\hline
			Setting&Speed&Memory&Smallest&Largest \\
			\hline
			Setting1 &[1, 180]&[1.5, 2]& HeteroFL&ResNet18\\
			Setting2 &[1, 180]&[1.5, 2]& FedRolex&ResNet18\\
			Setting3&[180, 360]&[2, 3]&ResNet18 & ResNet34\\
			Setting4&[180, 1K]&[2, 6]&ResNet18 &ResNet152 \\
			Setting5&[30, 100]&[1.5, 2]& ViT-tiny-tiny & ViT-tiny\\
			Setting6&[0,110]&\{4,8\}&BERT-tiny&BERT-large\\
			\hline
		\end{tabular}
	\end{center}
\end{table}

\subsubsection{Methods Implementation}
For fair comparison, implementation of baseline methods is the same as in their papers. In our methods and baselines, the sampled model should satisfy both the memory and bandwidth conditions. We construct our search space $(w, S)$ with $w$ as the biggest model in \cref{tab:model}. For the first two setting $S=\{0,0.0625, 0.125, 0.25,0.5,1\}$. For the third and forth setting $S=\{0,0.5,1\}$. For the fifth setting $S=\{0,1\}$. For the sixth setting $S=\{0,0.5,1\}$. With previous empirical observations, we set the frequency to check resource budgets at the communication round level instead of each local iteration level. When the client checks in FL, the resource budgets on it shall be stable enough to hold on for completing one communication round. 

\subsubsection{Algorithm Overhead}
Th overhead of our algorithm is only on the server as the clients only do the training of a single model the same as FedAvg. Comparing to the FedAvg, on the server, during each communication round, we will conduct resource constrained model search and $T_{SKD}\cdot n$ iterations of in-place distillation. Such processes will add little overhead. Our server has an NVIDIA RTX A4500 GPU and 12 CPU cores. From setting 1 to 5, 
all processes of resource constrained model search are completed less than 1 second in each communication round.  The latency spent on the server including model search and distillation in each communication round is 19.65, 15.64, 13.39, 19.16, and 20.5 seconds. Though the number of clients increases from setting 1 to 5, the overhead does not significantly increase. For setting 1, if we use FedAvg with the smallest model, it is 16.87s. The overhead will not increase when the number of clients increases, reflecting potential scalability.In usual case, the latency of federated learning is primarily occupied by the communication. Comparing to the turnaround time from the server sending models to receiving data from clients, which is usually over 100 seconds, the overhead of our algorithm is negligible.

\subsection{Comparison with  Baselines}
\begin{table}[tb]
	\caption{The inference accuracy of the global model under different settings. We use w/ and w/o to denote methods with in-place distillation.}
	\label{tab:results}
	\begin{center}
		\small
		\begin{tabular}{l@{\hskip 0.1in}c@{\hskip 0.14in}c@{\hskip 0.14in}c@{\hskip 0.14in}c@{\hskip 0.14in}c@{\hskip 0.14in}c}
			\hline
			Settings&  1 &2&3&4&5&6\\
			\hline
			\begin{tabular}[c]{@{}c@{}}FedAvg\\Largest \end{tabular}&71.91&75.74&92.14&74.51&84.39&65.38\\
			\hline
			\begin{tabular}[c]{@{}c@{}}FedAvg\\Smallest\end{tabular}&54.16&38.82&84.50 &71.91 &67.86&56.96\\
			\hline
			HeteroFL &61.64&63.90&88.76 &60.95&41.87&54.32\\
			FjORD&66.45&33.53 &88.61&47.17 &24.90&11.42\\
			\hline
			FedRolex &26.50&69.44&78.63 &30.53&9.22&22.39\\
			FedDropout &18.73&46.64&34.01 &14.99&22.64&14.43\\
			AnycostFL &14.87&14.96&87.75&69.08 &8.55&12.51\\
			\hline
			Ours (w/o) &69.02  & 70.74 &91.18&73.5&81.79&61.78\\
			Ours (w/)&\textbf{69.77}&\textbf{71.87} &\textbf{91.44} &\textbf{74.74}&\textbf{83.67}&\textbf{64.32}\\
			\hline
		\end{tabular}
	\end{center}
\end{table}

We conduct the comparison experiments under the six settings and the accuracy of converged global model is shown in \cref{tab:results}. For the method of FedAvg, we show the results of using the smallest and the biggest models. In the actual system, only FedAvg with the smallest model works with resource-constraints requirements on mobile devices. FjORD is a method combining system-heterogeneous federated learning and knowledge distillation. Our methods outperform existing system-heterogeneous FL methods in all settings with the benefits of adopting various model architectures and in-place distillation. In some cases, our methods can not only get close to the performance of adopting the biggest model with FedAvg, but better than that. This performance superiority exists in HeteroFL and FedRolex as well. It shows aggregating models of different architectures can help resolve data heterogeneity as well.

Lack of flexibility in existing methodologies is a significant factor contributing to the unsatisfactory outcomes in certain scenarios. In these cases, the range of resource constraints is broader, and more diverse resource constraints exist across mobile devices. For example, comparing setting 1 and setting 4 where the difference is that we set a more realistic resource constraints scheme, the performance of baseline HeteroFL and FedDropout will drop.  
For example, performance drops severely in particularly hard settings such as setting 4 and setting 5. For method AnycostFL, comparing setting 3, 4 with setting 1, 2, 5, when resources are more constrained, it is not a good idea to choose channels based on their L2 norms. Even on a smaller range of the heterogeneity over the resources in setting 1 and setting 2, we still can beat the baselines. These show that our methods can work in various settings of different scales of heterogeneity.

\begin{figure}[tb]
	\centering
	\begin{minipage}{0.48\linewidth}
		\includegraphics[width=\linewidth]{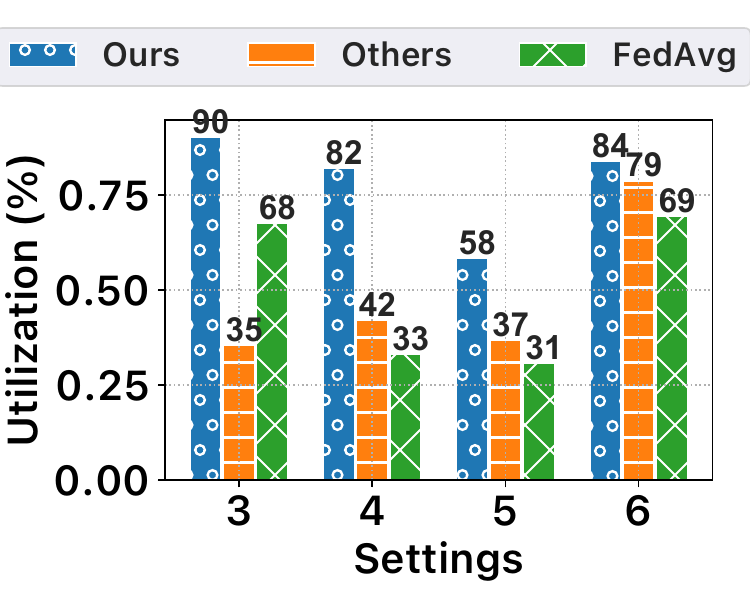}
	\end{minipage}\hfill
	\begin{minipage}{0.48\linewidth}
		\includegraphics[width=\linewidth]{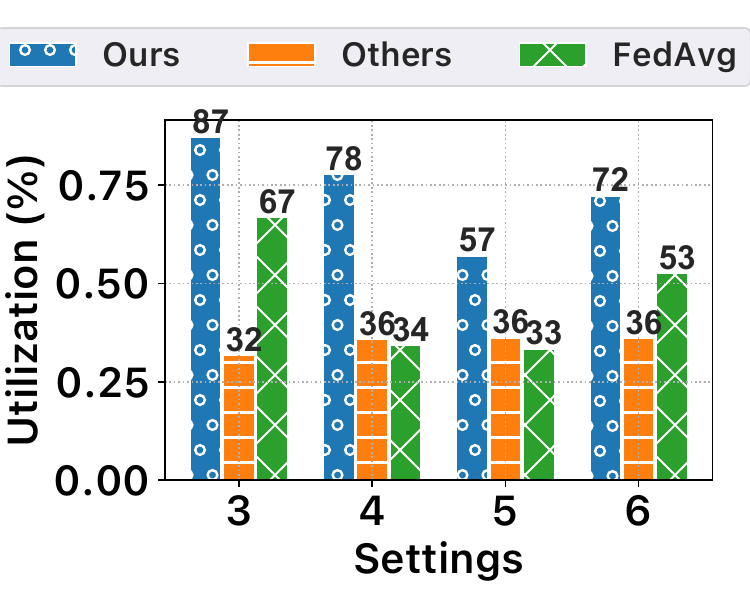}
	\end{minipage}
	\caption{The average utilization of GPU memory and bandwidth of different methods under different settings.}
	\label{fig:utilization}
\end{figure}

Another important metric is how well they fit into clients' system budgets such as the utilization of memory and bandwidth. We measure and compare the utilization of GPU memory and network bandwidth in \cref{fig:utilization}. Because previous methods have the same patterns of choosing only one fixed pruning rate for each model during the training process, they have the same utilization rate. However, this approach often results in sub-optimal model configurations, requiring higher pruning rates to meet both accuracy and utilization requirements, thereby diminishing overall performance.

\subsection{Effects of Search Space Design and Resource Constrained Search}
To delve deeper into the design of search space and effectiveness of resource constrained search . We conduct an experiment on setting 3 and the $S$ of the search space is $(\{1/16, 1/8, 1/4, 1/2, 1\})$, which is the same as the settings of five complexities in our baselines. In this case, we only have search dimensions of different widths on each layer, we still can have benefit comparing to applying one same width pruning rate for all layers, where it can achieve accuracy of $90.04\%$.

Apart from the search dimension of widths, another search dimension is depth. In the setting 3, we assign a group of experiment that the $S$ in the search space is \{0, 1\} and the variance of different architectures only lies in different depths. The accuracy is $90.3\%$. Our method can reach the accuracy very close to the accuracy acquired by the biggest model. In all the settings, even if the whole biggest model is never sampled to the client, we can still compose the biggest model. Taking the chance of small search space, we repeat the experiments by replacing resource constrained search with enumerating all architectures, the accuracy is 91.11\%. Hence, our resource constrained search can achieve the approximately optimal results.

\subsection{Effects of In-place Distillation}

To evaluate the effectiveness of our proposed federated in-place distillation module. We conduct experiments of removing it. From \cref{tab:results}, we can see under different settings, in-place distillation module can help improve the final accuracy. Notably, our method can still reach higher accuracy than baselines even without in-place distillation. Different from sBN module in HeteroFL~\cite{heterofl} and FedRolex~\cite{fedrolex} which may leak privacy as their server needs to query clients about batch normalization statistics, our in-place distillation does not leak privacy.

\begin{figure}[tb]
	\centering
	\begin{minipage}{0.48\linewidth}
		\includegraphics[width=\linewidth]{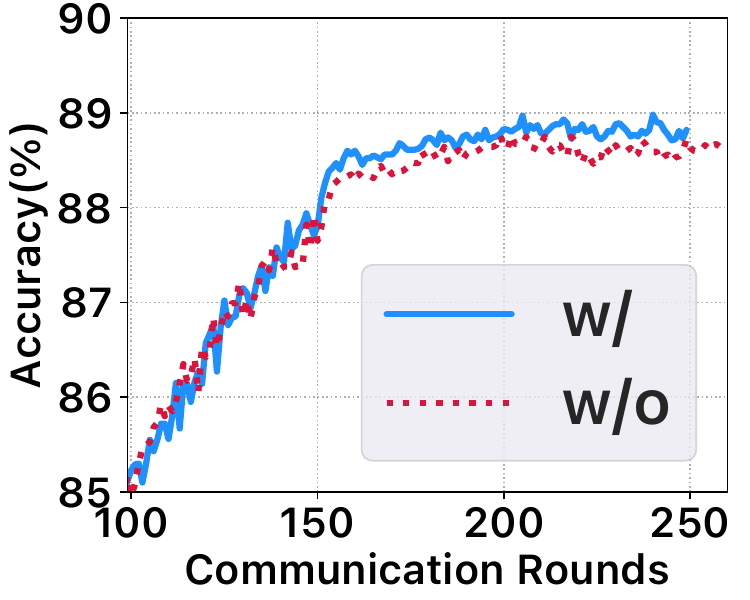}
	\end{minipage}
	\begin{minipage}{0.48\linewidth}
		\includegraphics[width=\linewidth]{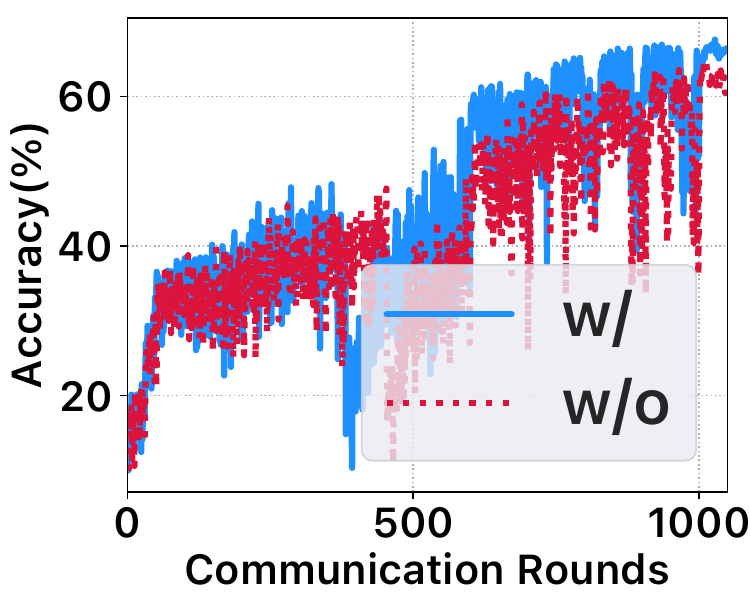}
	\end{minipage}
	\begin{minipage}{0.48\linewidth}
		\includegraphics[width=\linewidth]{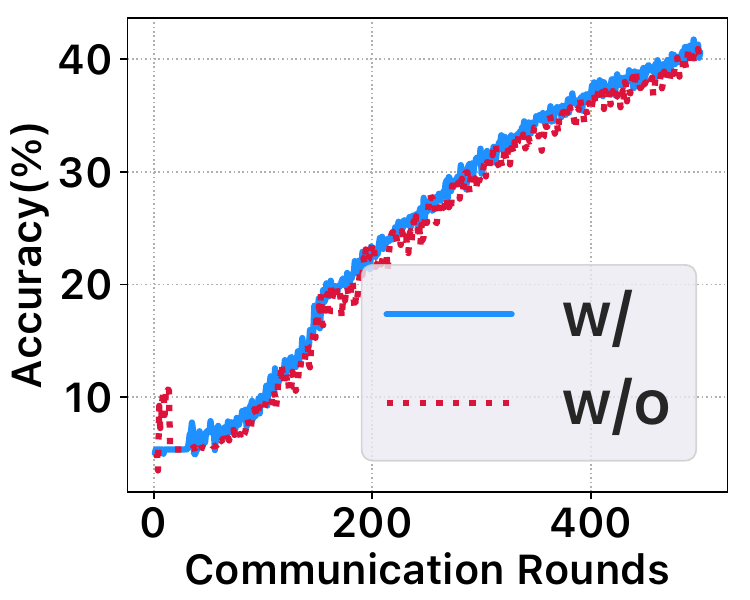}
	\end{minipage}
	\begin{minipage}{0.48\linewidth}
		\includegraphics[width=\linewidth]{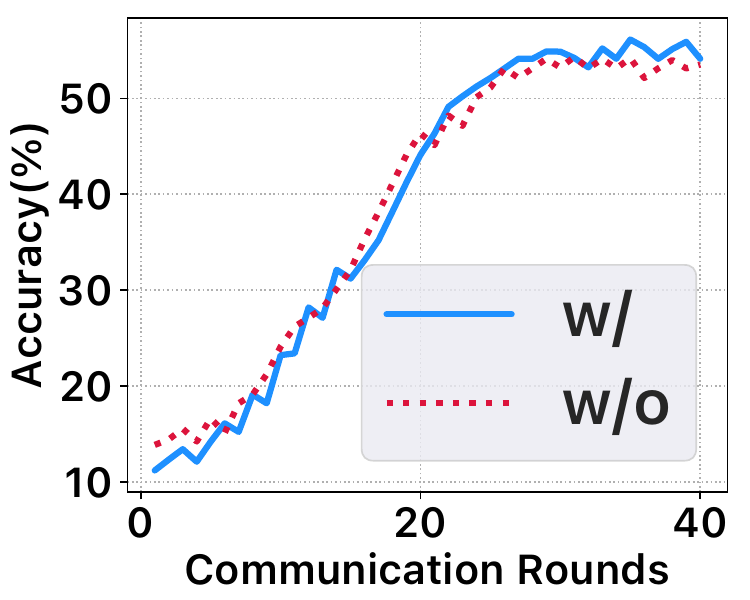}
	\end{minipage}
	\caption{The training curve of the test accuracy of the global model of HeteroFL with (w/) and without (w/o) inplace distillation from setting 3 to setting 6. }
	\label{fig:distill}
\end{figure}

On the other hand, since our module can be applied to other methods, we implement the process of in-place distillation to HeteroFL and the accuracy and convergence can be improved as shown in \cref{fig:distill}. The accuracy over HeteroFL can be improved to 88.98\%, 67.60\%, 41.73\%, and 56.13\% respectively from setting 3 to setting 6. Regarding FjORD self-distillation module, from \cref{tab:results}, we can see that it is helpful over i.i.d.~datasets (setting 3) but performs poorly over non-i.i.d~case (setting 2, 5, 6). Because they conducted distillation over the clients' datasets. Such biased distillation on clients can lead to over-fitting on clients' datasets and poor performance. In contrast, our method can provide performance improvement in both i.i.d~and non i.i.d.~datasets. Apart from this, though we both share the same extra iterations of knowledge distillation, our in-place distillation is conducted on the server while FjORD is on the clients. As the server usually has more powerful computation ability, our method is more efficient and client-friendly.

	\section{Conclusion}
We revisit the system heterogeneity problem in federated learning with a new system which can compare algorithms on board with logs collected on mobile devices. We focus on the problem of how we can optimize the models assigned to each client and utilize resource budgets (memory and bandwidth) on mobile devices to the maximum. Existing system-heterogeneous methods typically use channel width pruning with a fixed prune rate for all layers of a neural network, which will lead to under-utilization of client available resources and poor performance. To address this issue, we propose assigning models of various architectures to clients, which allows for greater flexibility and better resource utilization. To efficiently sample these models, we propose a resource constrained model search, and introduce federated in-place distillation to improve performance. Such in-place distillation is applicable to existing system-heterogeneous methods as well. We demonstrate the effectiveness of our method in various  settings, highlighting its potential to enhance system efficiency and client satisfaction.
	
	\newpage
	\section*{Broader Impact}
	This paper presents work whose goal is to advance the field of Machine Learning. There are many potential societal consequences of our work, none which we feel must be specifically highlighted here.
	\bibliographystyle{icml2024}
	\bibliography{main}
	\newpage
	
	\newpage
	\appendix
\section{Additional explanation of Overall Algorithm}
\label{appendix:overallalg}
\begin{algorithm}[tb]
	\caption{System-heterogeneous federated learning through dynamic model search}
	\label{alg:fl}
	\begin{algorithmic}[1]
		\STATE {\bfseries Input:} $N$ clients and the search space $(w,S),\epsilon,T_{\rm MAX}$.
		\STATE Initialize the global model $w_0$.\;
		\STATE Initialize the sampling pool: 
		\\$\mathcal{P}=\{(s_1,\cdots,s_1), (s_m,\cdots,s_m)\}$;
		\FOR {each round $t\leftarrow1$ to $ T$}
		\STATE {\bfseries On Server: }\;
		\STATE Sample $N'$ clients: $\mathcal{N'}\subseteq \{1,\cdots, N\}$.\;
		\FOR {Each client $i\in \mathcal{N'}$}
		\STATE Binary search and sample $V_i$ from $\mathcal{P}$ with condition $b_i,c_i$.\;
		\STATE ${\rm loop}\leftarrow 0$.\;
		\WHILE{probability bigger than $\epsilon$}
		\STATE Random search a new $V'$ from $(w,S)$.\;
		\STATE $\mathcal{P}\leftarrow \mathcal{P} \cap\{V'\}$.\;
		\IF{$V'$ is better than $V_i$}
		\STATE $V_i\leftarrow V_i'$.\;
		\ENDIF
		\STATE ${\rm loop}\leftarrow {\rm loop+1}$.\;
		\IF{${\rm loop}>T_{\rm MAX}$}
		\STATE Break.\;
		\ENDIF
		\ENDWHILE
		\STATE Sample client model $w_i=w\otimes V_i$ from the global model and then send it to client $i$.\;
		\ENDFOR
		\STATE {\bfseries On Client: }
		\FOR {on client $i\in \mathcal{N'}$ parallel}
		\STATE Conduct local updates.
		\STATE Upload $w_i$ and current constraints $(b_i,c_i)$ to the server.\;
		\ENDFOR
		\STATE {\bfseries On Server: }
		\STATE Receive $w_i$ and aggregate $w_i$ to update $w$.\;
		\STATE Conduct process of inplace distillation.\;
		\ENDFOR
	\end{algorithmic}
\end{algorithm}

The whole process of system-heterogeneous federated learning through dynamic model search including the algorithm of resource constrained model search (from line 7 to line 19) is shown in the \cref{alg:fl}. 

\section{Examples of Choosing Different Channels on Each Layer}
\begin{figure}[tb]
	\centering
	\includegraphics[width=0.8\linewidth]{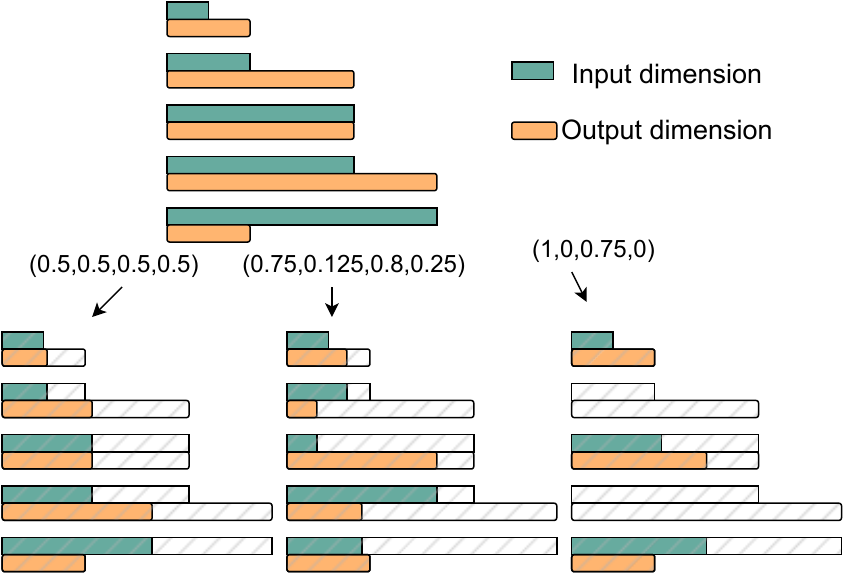}
	\caption{The illustration of how we can implement the models assigned to the client according to definition ~\ref{def:subnet}.}
	\label{fig:supernet}
\end{figure}

In \cref{fig:supernet}, we give three examples. The input channel of the first layer depends on the input and the output channel of the last layer depends on the output. The output channel of each layer is decided by $v_i$ and the input channel of each layer is decided by $v_{i-1}$. For the first example, it is the same as the model generated in previous methods where $v_i$ are all the same. For the second example, each layer has a different prune rate. For the third example, we can change the depth of the model, and each layer is removable.

\section{Additional Explanation about Federated In-place Distillation}
\label{appendix:distill}
We derive \cref{eqa:skd} from \cref{eqa:kd}. As on the server, we do not have the datasets, we only consider the second part in \cref{eqa:skd}. To update the supernet, we use the gradient:
\begin{align}
	\nabla w&=\mathbb{E}_{w_i}\mathcal{L}_{\rm KD}([w,w_i])\\
	&=\frac{1}{n}\Sigma_{i\in[n]}\mathcal{L}_{\rm KD}(w,w_i;\mathcal{X_{\rm KD}})
\end{align}
The main idea behind our in-place distillation module is that from \cref{eqa:skd}, we can minimize the distribution divergence between the global model and its subnets through minimizing the KD loss. In such a way, the sampled networks will have a closer output distribution to the global model and so that the aggregation process will be more effective and the performance of the global model will be improved. 

During the process of training, the global model is trained in two steps iteratively. It is first trained with aggregation from client models and then trained through in-place distillation. Such a process is repeated through several communication rounds. So, we can estimate the loss of the global model:
\begin{align}
	\mathcal{L}(w)&=\eta \frac{1}{N}\Sigma_{i\in[N]}\mathbb{E}_{(x,y)\sim D_i}\left[ \mathcal{L}_{\rm obj}(y,w_i(x))\right]\\&+\gamma \frac{1}{n}\Sigma_{i\in[n]}\mathbb{E}_{(x,y)\sim N(\mu_\mathcal{D},\sigma_\mathcal{D})}\left[ D_{\rm KL}(w(x)\Vert w_i(x))\right]
\end{align}
where $\eta$ and $\gamma$ are estimated learning rates for these two steps. The first part of the loss function is an estimation of the expectation and is actually realized through aggregation of client models as the global model is not really trained. The global model is bounded by the objective of federated learning and inplace distillation to minimize the distribution drifts between different clients. As a result, our global model can be viewed as optimized through two targets and thus the performance can be improved. 
\section{Proof of Convergence Analysis}
\label{appendix:convergence}
Here we give the proof of \cref{theorem:convergence}.
\begin{proof}
	The key guarantee of the convergence is the boundary over the divergence of $w_i$s, we assume in the FedAvg 
	\begin{equation}
		\mathbb{E}\left[ \sum_{i\in N'}p_i \lVert w^t-\upsilon_i^t\rVert^2\right]\leq G
	\end{equation}
	where $G$ is the boundary, $\upsilon_i$ are client models which has the same structure as the global in FedAvg and $p_i=\frac{\lvert D_i\rvert}{\sum_{j\in[N']}\lvert D_j\rvert}$. 
	During the model search step, we sample the $w_i$ which is a subnet from the global model. For the parameters which are not sampled in $w_i$, they will be 0. As a result, we can construct a virtual model $\theta_i^t$.  This virtual model has the same architecture as the $w^t$ but the values of the parameters are different. The parameters should be the same as those in $w_i$ if they are not 0. For the rest parameters, we can assign them particular values to make them have the same distribution as $\upsilon_i^t$. We will have
	\begin{equation}
		\mathbb{E}\left[ \sum_{i\in N'}p_i \lVert w^t-w_i^t\rVert^2\right]\leq \mathbb{E}\left[ \sum_{i\in N'}p_i \lVert w^t-\theta_i^t\rVert^2\right]\leq G
	\end{equation}
	According to the convergence analysis of federated learning~\cite{convergence}, the convergence rate is $\mathcal{O}(\frac{1}{T})$.
\end{proof}

\section{Supplementary for System Design}
\subsection{Server Module}
On the server, we will import the user configurations and the server will correspondingly assign the resource budgets. We show the example configurations in \cref{fig:configuration}. In the \emph{limitation} parameter, users will first list out the types of resources, for example, memory, bandwidth, energy. Under each kind of resource, users have two choices. They can either assign a maximum and minimal value or assign a path to load logs. If users choose the former, the budget will be uniformly sampled from the corresponding range. In the former option, if the user sets binary to true, the value will be only sampled from the two values (maximum and minimum). If they choose the latter, the server will import logs of each client device. Each log file should log the timestamps and corresponding values. In \textsc{Plato}, as the wall clock elapsed time is provided, we can quickly search the corresponding resource budget at the timestamp we need to assign models.

We also provide an interface for checking the resource usage of a given model structure in the server. When the algorithm module searches for the proper model structure, the server will evaluate how much memory it will use or how much time it will take to transmit the model according to the current set of resource budgets.
\subsection{Client Module}
The client module is composed of a resource monitor and a \emph{Trainer} module. The trainer will receive the assigned model and train with local data. For the resource monitor, it will keep a track of current resource utilization, If current utilization exceeds the assigned budget, the client will throw corresponding errors. Besides this, we may also consider the case that during local training, the available budgets can change, which though happens scarcely. Hence, we will upload the client logging to the server module. The server module will compare the client logging, local training time and logging file in the user configuration file. If during client training, the resource exceeds the budget, the server will treat this time to training on the client as a failure.

\section{Details about Datasets and Models}
\label{sec:datasets}
FEMNIST needs 3597 clients in total and each client has  $226.8\pm88.94$ samples. Shakespeare needs 1129 clients and each client has $3743.2\pm6212.26$ samples. For CIFAR10, training samples are equally partitioned among clients. For the model families, we have ResNet~\cite{resnet}, vision image transformers~\cite{vit} for image classification and BERT~\cite{bert} for next character prediction. For the first two settings, the smallest model is the smallest model defined in the corresponding papers: ResNet18 with channels pruned to 6.25\%. ViT-tiny-tiny is the ViT-tiny with only one depth. BERT large is the same as the model in the paper~\cite{bert} and BERT-tiny only has 6 layers, hidden dimension 256 and attention head 6. The resolution of the images is $32$. The sequence length of the text task is 80. 

\section{Hyper-parameter Settings}
\label{sec:hyper}
Here we show the rest settings of the hyper-parameters during experiments. For the setting1, 3, and 4, the hyper-parameters are the same as parameters used in HeteroFL. For the setting 2, they are the same as those in FedRolex. For setting 5, they are the same as hyper-parameters in training ViT on ImageNet~\cite{vit}. For setting 6, they are the same as hyper-parameters in training BERT~\cite{bert}. For the setting 1,3, 4 and 6, we use the SGD optimizer with learning rate 0.1, momentum 0.9 and weight decay 0.0005. For the setting2, the learning rate is 0.0002 and the rests are the same. For the setting 5, we use AdamW optimizer with learning rate 0.001, $\beta$s are $(0.9,0.99)$ and weight decay is 0.01. We also use the learning rate scheduler, for the setting 1 to 4 and 6, we use the multi-step scheduler with a decay rate of 0.1. For setting 1, we decay at round 300 and 500. For setting 2, we decay at round 800 and 1250. For setting 3, we decay at 150 and 250. For setting 4, we decay at 300 and 500. For the setting 6, we decay at 10 and 20. For setting 5, we use the cosine scheduler and the cycle is 500 rounds.

\subsection{Hyper-parameters in Resource Constrained Model Search}
During the resource constrained model search, there are two important hyper-parameters: $T{\max}$ and $\epsilon$. Intuitively, larger $\epsilon$ will lead to less exploration,  which means less possibility to find proper structure.  On the other hand, more exploration involves more random search, which means possible more running time. Larger $T_{\max}$ will increase the possibility of expanding the sampling pool and find more suitable structures but can lead to increasing running time. Before start the process of federated learning to measure the accuracy and efficiency, we do an experiment to find out the suitable $\epsilon$ and $T_{\max}$.

\begin{figure}[tb]
	\centering
	\includegraphics[width=0.95\linewidth]{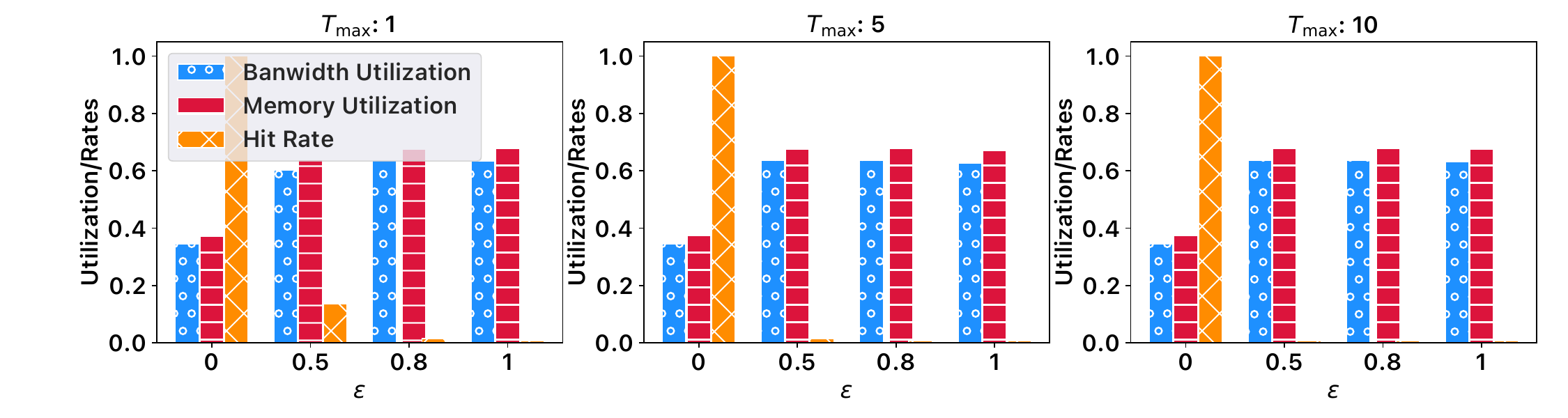}
	\caption{The utilization of bandwidth and memory as well as hit rates with different $\epsilon$ and $T_{\max}$ during resource constrained model search.}
	\label{fig: debug}
\end{figure}

In this experiment, we do not really train each model but only measure the utilization of assigned GPU memory and bandwidth with the given single model on each client. Higher utilization means we find a more suitable model to fit into the given GPU memory and bandwidth and such a model can better help train the global model. We also measure the rate that whether the $T_{\max}$ is reached, which we define as hit rates. We use setting 5, which uses the logs from the real world and simulates for 100 communication rounds. \cref{fig: debug} shows the average utilization and hit rates among all participant clients and communication rounds. We can see that lager $\epsilon$ can help get better utilization. To introduce the exploration into the search process, we adopt $\epsilon=0.8$. We can see that when $T_{\max}$ increases, the hit rate will be smaller. However, too large $T_{\max}$ may lead to worse efficiency. So we adopt $T_{\max}=5$.

\subsection{Hyper-parameters in In-place Distillation}
There are two hyper-parameters we need to set during the in-place distillation. We use the Adam optimizer with learning rate 0.001, which is a common setting of the optimizer. The $T_{\rm SKD}$ is set as 100. There is no particular value for this hyper-parameter. Any reasonable value is fine. We find that after $T_{rm SKD}$, it does not bring much accuracy improvement in the final results. But the latency spent on the server will linearly grow. So we choose the value of 100.

\section{Details of 4G/LTE logs}
\begin{figure}[tb]
	\centering
	\begin{minipage}{0.49\linewidth}
		\includegraphics[width=\linewidth]{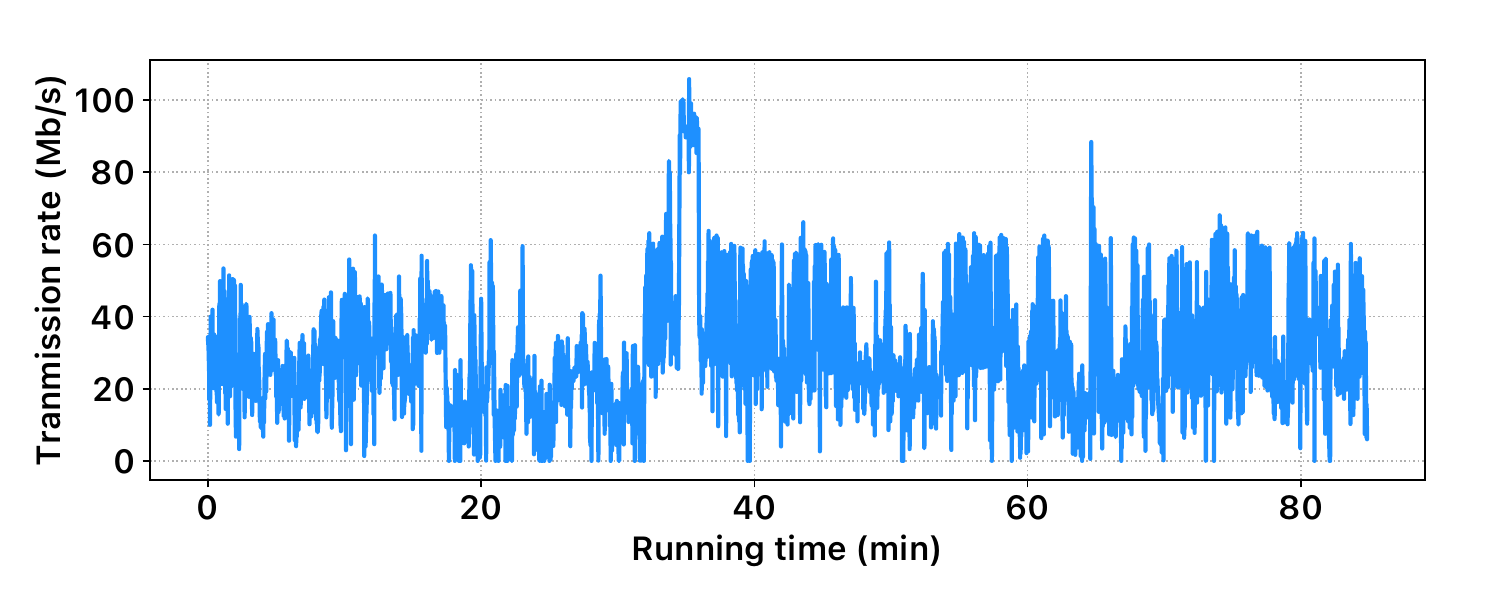}
	\end{minipage}\hfill
	\begin{minipage}{0.49\linewidth}
		\includegraphics[width=\linewidth]{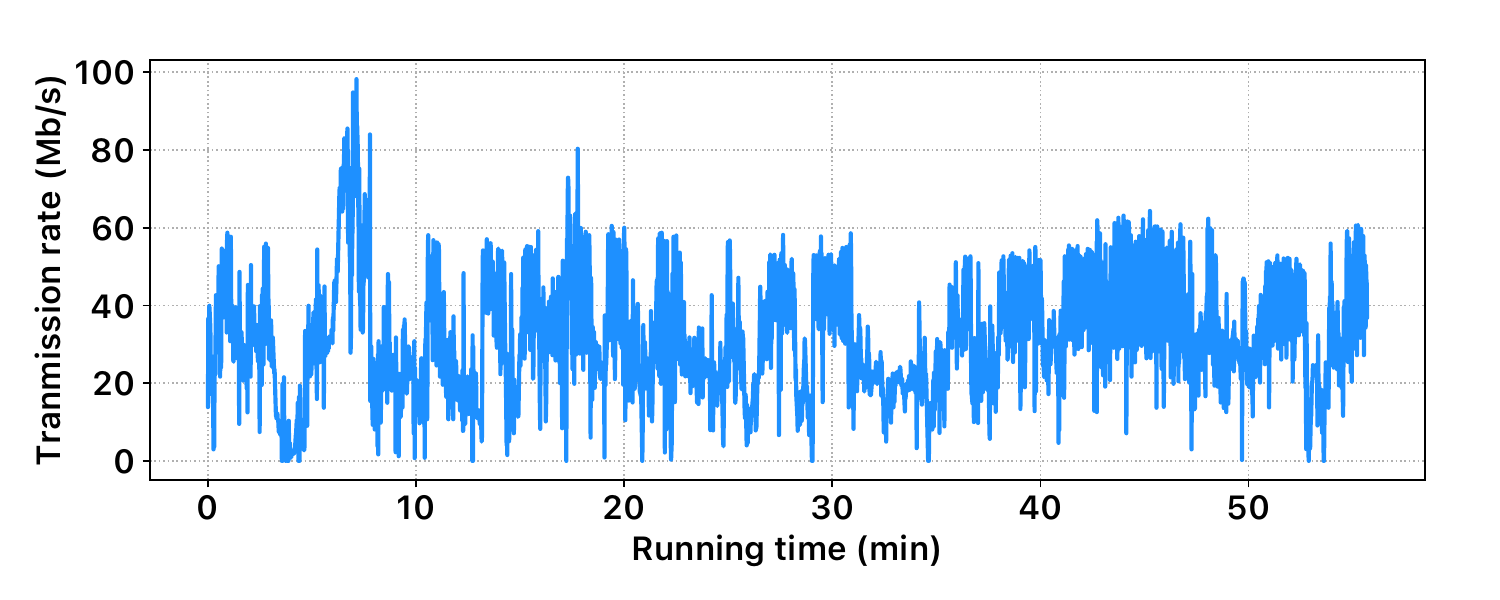}
	\end{minipage}\hfill
	\begin{minipage}{0.49\linewidth}
		\includegraphics[width=\linewidth]{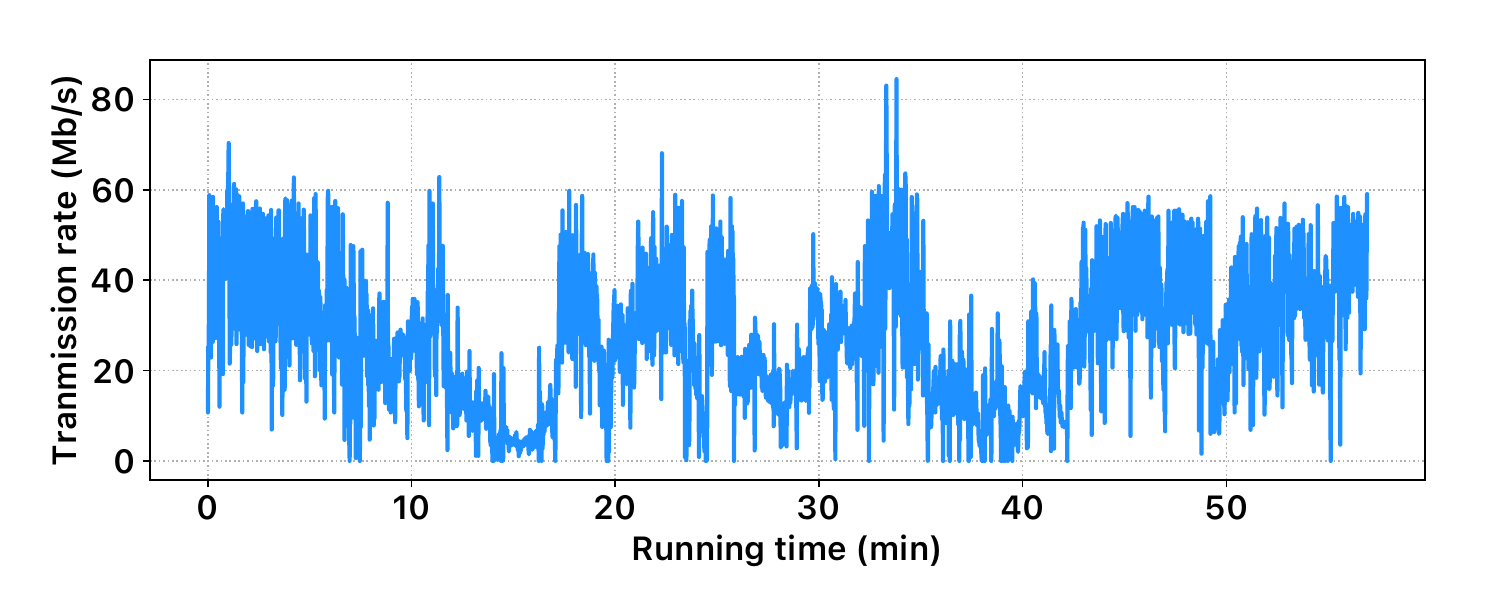}
	\end{minipage}\hfill
	\begin{minipage}{0.49\linewidth}
		\includegraphics[width=\linewidth]{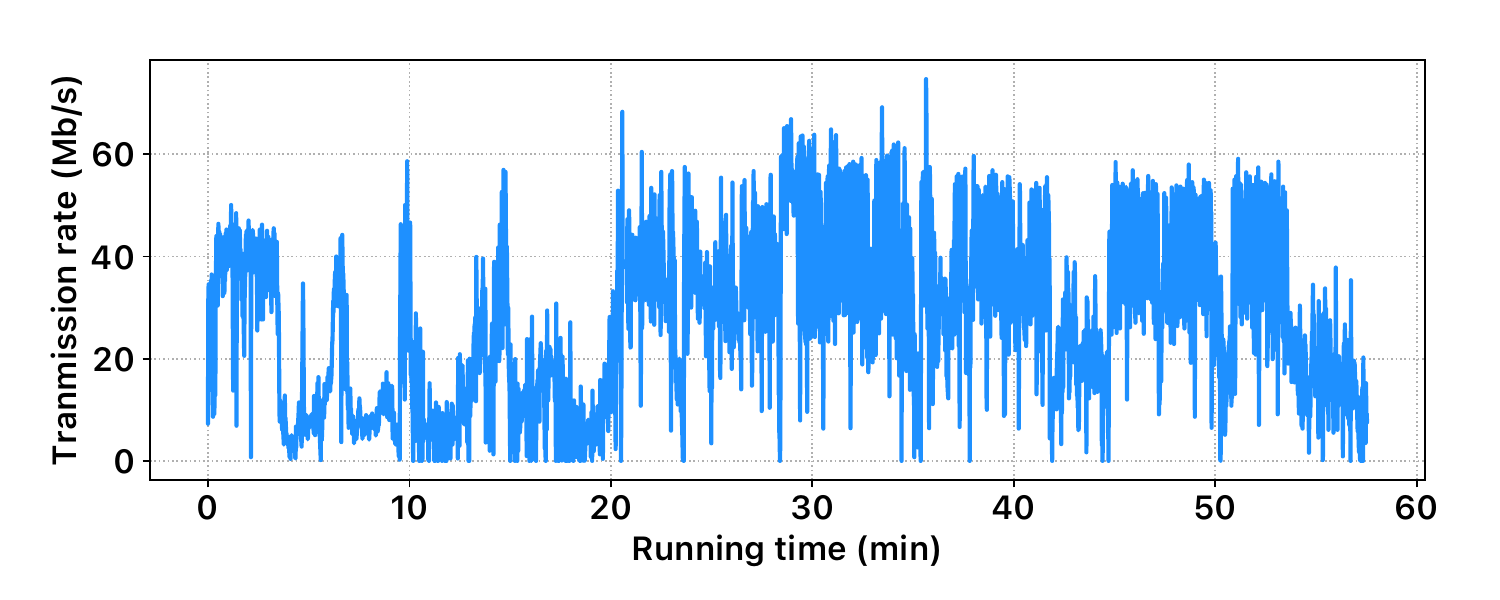}
	\end{minipage}\hfill
	\caption{The visualization of 4G/LTE bandwidth logs over different public transportation: Bus, Car, Foot, and Tram}
	\label{fig:logs}
\end{figure}

We show the details of these logs in \cref{fig:logs}.

\section{Additional experiments}

\begin{figure}[tb]
	\centering
	\begin{minipage}{0.33\linewidth}
		\includegraphics[width=\linewidth]{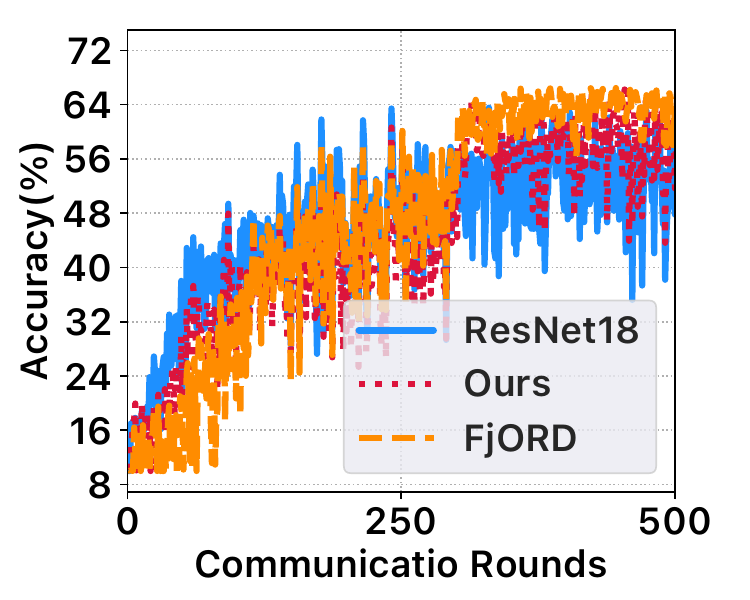}
	\end{minipage}\hfill
	\begin{minipage}{0.33\linewidth}
		\includegraphics[width=\linewidth]{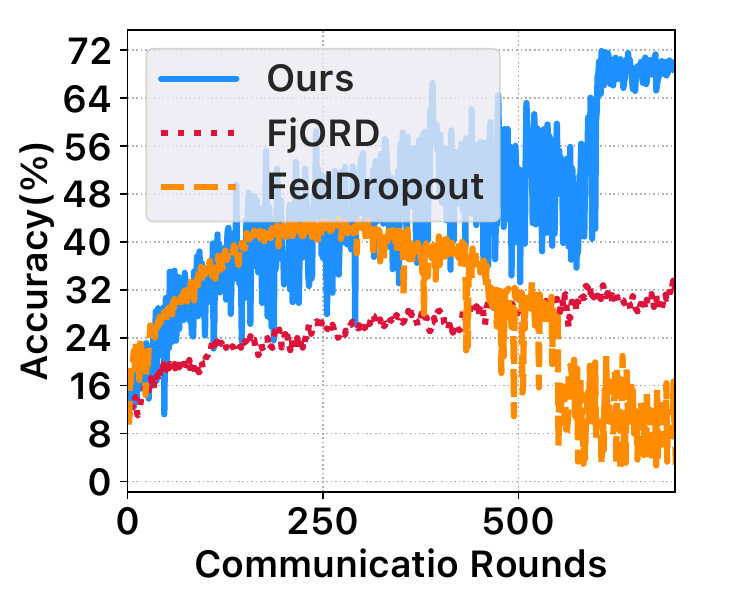}
	\end{minipage}\hfill
	\begin{minipage}{0.33\linewidth}
		\includegraphics[width=\linewidth]{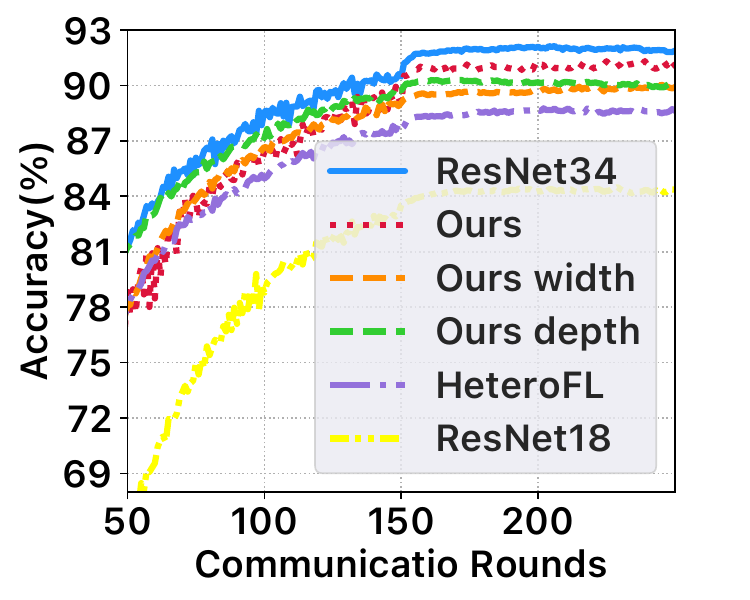}
	\end{minipage}\hfill
	\begin{minipage}{0.33\linewidth}
		\includegraphics[width=\linewidth]{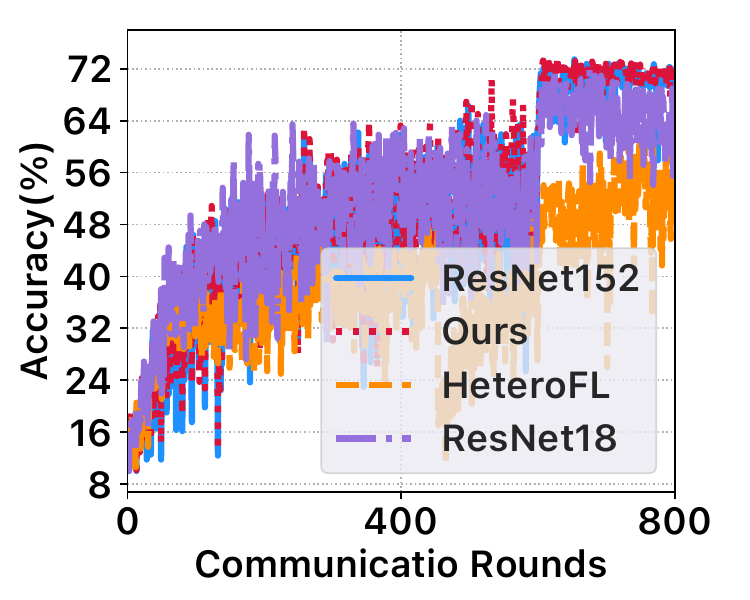}
	\end{minipage}\hfill
	\begin{minipage}{0.33\linewidth}
		\includegraphics[width=\linewidth]{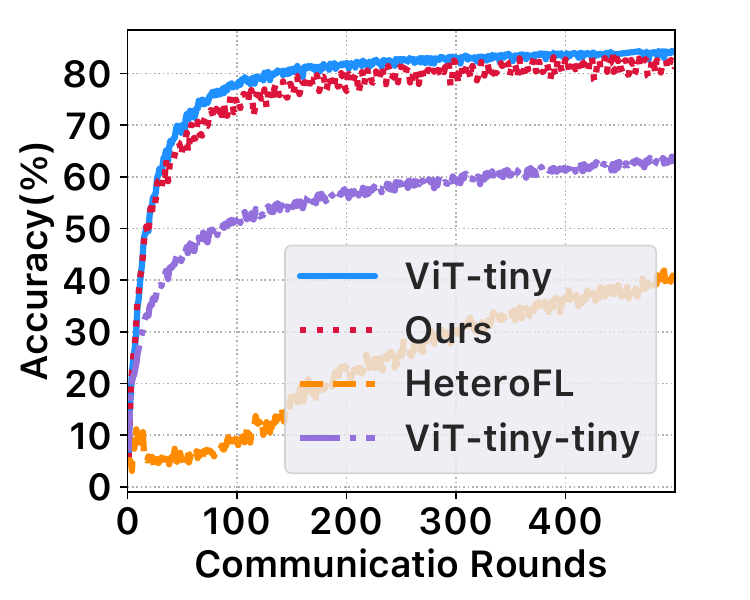}
	\end{minipage}\hfill
	\begin{minipage}{0.33\linewidth}
		\includegraphics[width=\linewidth]{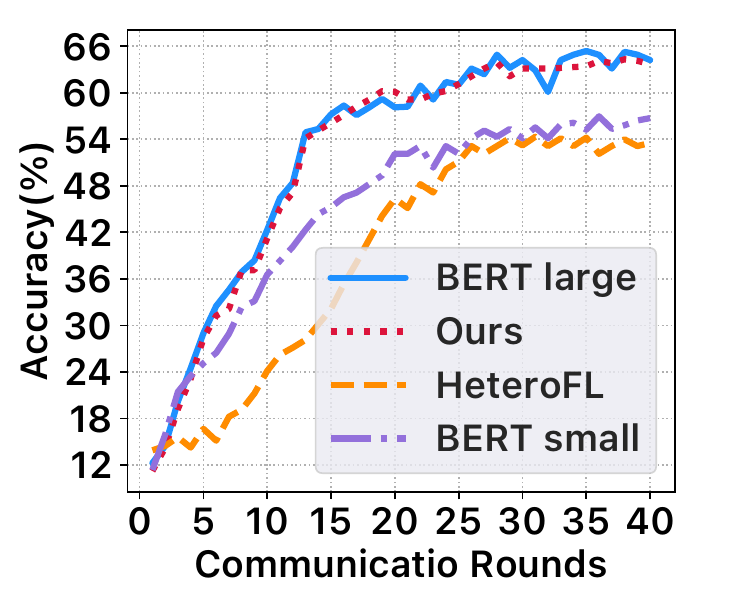}
	\end{minipage}\hfill
	\caption{The training curve of the test accuracy of the global model of our methods and the baselines under from setting 1 to setting 6. Ours width/depth means our method with search space only containing different widths/depths.}
	\label{fig:convergencecurve}
\end{figure}

\subsection{Comparison with Baselines}
The convergence rate during the training process is shown in \cref{fig:convergencecurve}.
\begin{figure}
	\centering
	\includegraphics[width=0.33\linewidth]{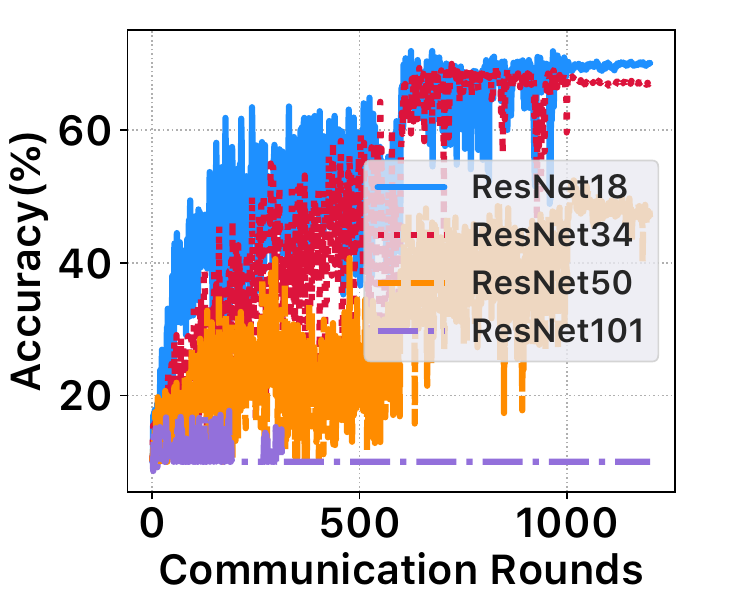}\hfill
	\includegraphics[width=0.33\linewidth]{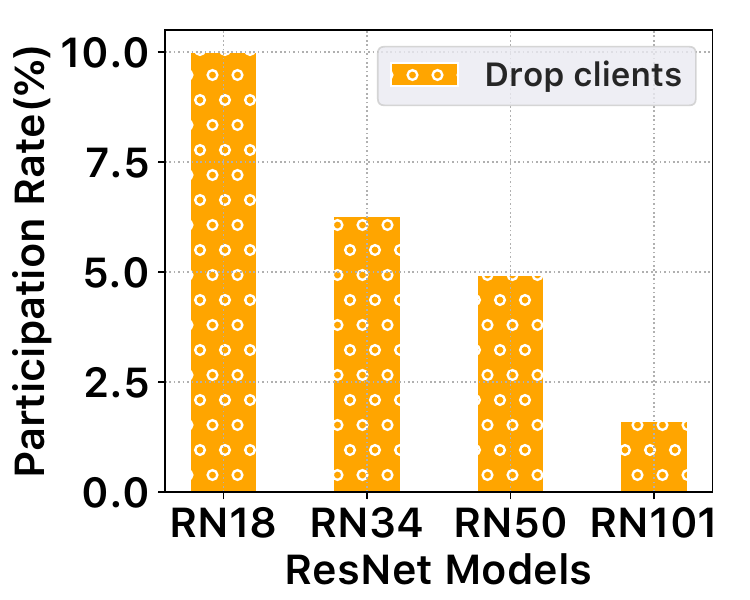}\hfill
	\includegraphics[width=0.33\linewidth]{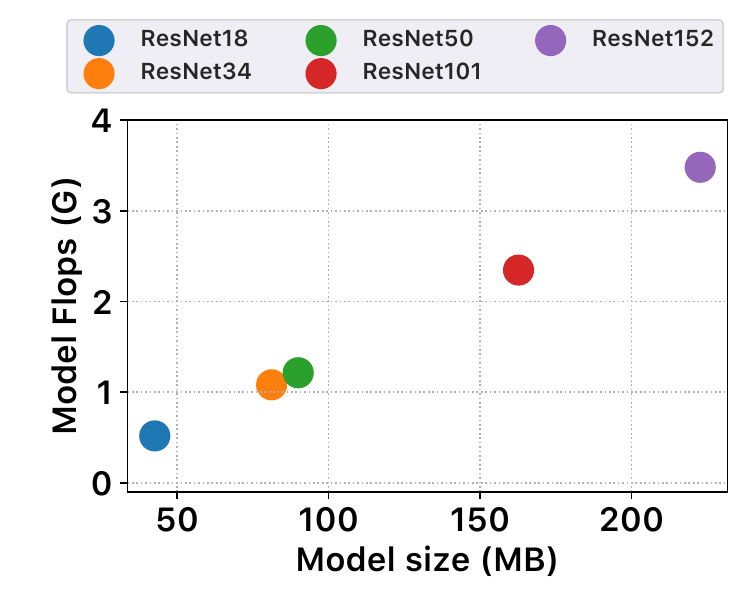}
	\caption{The global model accuracy and client participation of federated learning with clients dropping due to clients' various resource constraints. }
	\label{fig:Drop}
\end{figure}
\subsection{Impact of Failing to Meet Resource Constraints}
\label{subsec:drop}
In federated learning, when facing resource constraints, a trivial solution is to employ a smaller model. However, this may result in lower accuracy than what can be achieved with larger models. Given that resource constraints vary among clients in FL, some clients may possess unused resources that could be leveraged to train larger models. 

Another solution is to exclude low-resource clients. Nonetheless, if these clients persistently have limited resources, they will never be able to participate in federated learning. We conducted toy experiments (Non IID $\alpha=0.1$, 10/100) where we have 100 clients and 10 out of them are selected in each round. We construct a non-i.i.d.~CIFAR10~\cite{cifar10} dataset with Dirichlet distribution (parameter $\alpha=0.1$). The local epoch in each round is 5. We will exclude clients if they cannot run the model after client selection.  The resources limitation on each client is fixed and will not change between each communication round.

We choose different ResNet models leading to different participation rates. The results as well as the sizes and flops of these models are shown in \cref{fig:Drop}. Lower participation rate will result in low performance because we cannot make use of data on most of the clients.  In this case, this solution is even worse than adopting a smaller model. As a result, we need a resource-aware solution to have a high utilization of available resources, achieving high accuracy. 

Rather than adopting uniform models, there are approaches assign heterogeneous models to clients. Numerous existing methods focus on optimizing communication and computation overhead on clients, such as structured pruning, compression, and quantization. However, untargeted optimization cannot solve the hard-constraints problem. For instance, Hermes~\cite{hermes} uses structured pruning to balance communication and computation efficiency. It prunes channels with the lowest magnitudes in each local model and adjusts the pruning amount based on each local model's test accuracy and its previous pruning amount. But its adaptive pruning rate can still result in client failure to run the model because the prune rate is not calculated directly from the clients' resource constraints. Additionally, at the early stage of the algorithm, the clients and the server must transmit the entire model. Hence, even with the implementation of heterogeneous models, issues in system-heterogeneous FL may not be entirely resolved.

\subsection{Stability and Convergence}
\begin{figure}[tb]
	\centering
	\begin{minipage}{0.47\linewidth}
		\includegraphics[width=\linewidth]{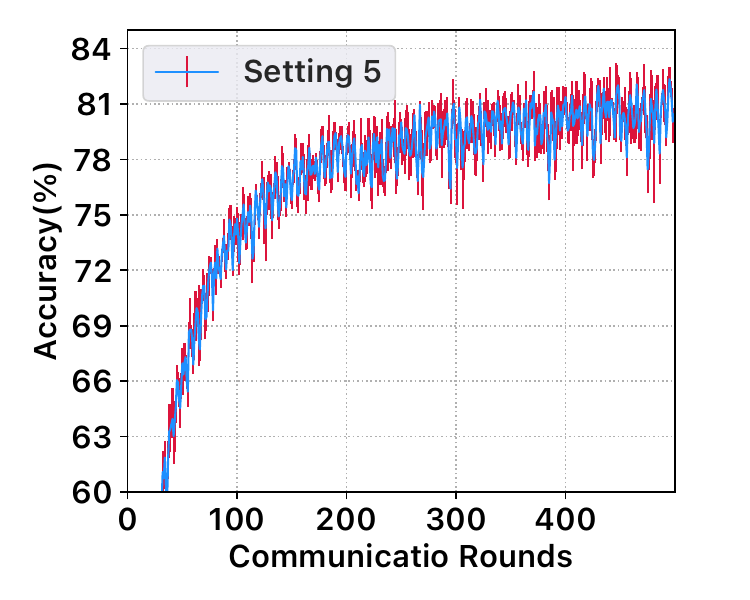}
		\caption{The training curve of the test accuracy by global model of our method with error bars under setting 5 using different random seeds.}
		\label{fig:stability}
	\end{minipage}\hfill
	\begin{minipage}{0.47\linewidth}
		\includegraphics[width=\linewidth]{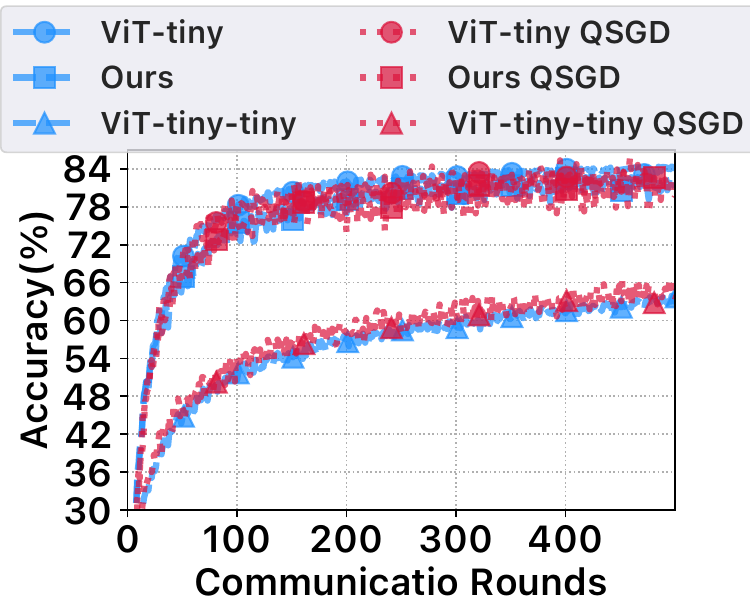}
		\caption{The training curve of the test accuracy by global model of our method and FedAvg after applying QSGD.}
		\label{fig:qsgd}
	\end{minipage}
\end{figure}

As there are randomization mechanisms in our methods during resource constrained model search and inplace distillation, we also test our methods over different random seeds to see if the randomness will affect the performance of our method. We repeat the experiments of setting5 with setting different seeds. The best inference accuracy of the global model in these three experiments are 83.20\%, 83.15\%, 83.67\%. As shown in \cref{fig:stability}, the convergence and final accuracy will not be affected by the randomness.

Besides theoretical convergence guarantee and analysis of our randomness effects on the communication round in \cref{subsec:converge}, we further show the convergence through empirical experiments. From the convergence curve in \cref{fig:convergencecurve}, our method can converge at the same rate or even faster than the baselines. In setting 1, our methods only need 311 rounds to reach the accuracy reported by HeteroFL where their setting for communication rounds is 800. In setting 2, our method only needs 1263 rounds to reach the accuracy reported by FedRolex where their setting for communication rounds is 2500. As a result, we are also efficient in total latency.

\subsection{Orthogonal to other Efficiency Methods}
We have shown that our methods can better solve resource-aware problems in federated learning compared to existing system heterogeneous methods. There are other conventional distributed learning efficiency methods. Though they cannot resolve system heterogeneous problems, they still can help improve efficiency. Our method is orthogonal to those methods such as compression and quantization. We apply the QSGD~\cite{qsgd} which is a compression method to reduce the communication overhead in distributed systems. The quantization level is 64 and we quantize from 32 bits to 8 bits. We apply this process to our methods and FedAvg and the results in shown in \cref{fig:qsgd}. Our method still works after applying QSGD. As a result, our method is orthogonal to previous efficiency methods and the combination of such methods with our methods will further help improve system efficiency.

\section{Discussion}
A limitation of our system is that we mainly develop the system with \textsc{Pytorch} framework. As a result, the interface for counting CUDA memory utilization are based on the \textsc{torch} API. As \textsc{Plato} also support other deep learning frameworks such as \textsc{Tensorflow} and \textsc{Mindspore}, we may further extend such interfaces of maintaining resource over other deep learning frameworks.

In this paper, we focus on two types of resources: memory usage and network speed. The reason is that these two kinds of resources may change more frequently than other possible resources during the federated training process, which can emphasize the importance of handing system-heterogeneity during federated learning. During federated learning, devices can also be made of hardware with different computation ability and energy consumption. However, users will not frequently change energy mode during the few hours learning process. The computation ability will not change during federated learning. So we place them in a minor consideration. While on the other hand, our system's modular design allows us to add in other constraints. In our algorithm, we can also add or change the resources with setting different constraints during the implementation of line 8 in \cref{alg:fl}. In short, our system and proposed model search method can have an easy extension to other resources besides memory and bandwidth.

Another direction of future work is about privacy. In this paper, in the perspective of each client, the training process the same as that in FedAvg. In other words, methods which are applicable to conventional federated learning such as differential privacy can be also applicable to our algorithms. Our system is developed upon Plato, where a lot of privacy-preserving methods in federated learning are built-in. However, an open challenge is whether aggregation of different architectures will affect privacy protection. As this paper focuses on system level optimization, we leave this for future work.

In our system, we have the flexibility of choosing logs for users. Users may need to ensure the accuracy of logs before inputting them into the system. The problem we are trying to resolve is allowing users to compare different system heterogeneity fairly with reproducible logs. Users can use any logs they want, the real-world logs, the logs generated by simulators or even logs under unreal settings. For example, the settings in HeteroFL and FedRolex cannot reflect practical use cases, but we can still compare different methods using their settings under our system fairly.

To resolve the question of how we can utilize available memory and network bandwidth to the maximum, besides system-heterogeneous federated learning, another direction is leveraging split learning. However , split learning requires the clients and the server to communicate at every iteration, which is quite inefficient in communication, compared to federated learning. Apart from that, to ensure privacy is preserved in split-learning, users still need to load certain layers on the clients. In other words, if users put a few layers on the clients to meet resource budgets, privacy may not probably be well preserved.
\end{document}